\documentclass[twocolumn,showpacs,showkeys,prl,superscriptaddress]{revtex4-2}
\usepackage{amsmath,amssymb,amsfonts,dsfont,mathrsfs,amsthm}
\usepackage{graphicx}
\usepackage{centernot}
\usepackage[justification=justified,singlelinecheck=false]{caption}
\usepackage{xcolor}
\usepackage{natbib}
\usepackage{url}
\usepackage{hyperref}
\usepackage{subcaption}

\hypersetup{
    colorlinks=true,
    urlcolor=black,
    linkcolor=black,
    citecolor=black,
    linktocpage,
    pdfborder={0 0 0}
}

\usepackage{feynmf}
\usepackage{siunitx}
\usepackage{array}
\usepackage{ulem}
\usepackage{tikz}
\usepackage{braket}
\usepackage{epsfig}
\usepackage{epstopdf}
\usepackage{tabularx}
\usepackage{float}
\usepackage{ragged2e}
\usepackage{booktabs} 

\usepackage{etoolbox}
\begin{document}


\title{Power-Law Suppression of Superfluid Stiffness in High-Kinetic-Inductance NbN Films}

\author{Meenakshi Sharma $^{\dagger\dagger}$}
\email{meenakshi.sharma@cpfs.mpg.de, $^\dagger$ uri.vool@cpfs.mpg.de}
\thanks{$^{\dagger\dagger}$These two authors contributed equally to this work.}
\affiliation{Max Planck Institute for Chemical Physics of Solids, Nöthnitzer Str. 40, 01187 Dresden, Germany}
\affiliation{Leibniz Institute for Solid State and Materials Research Dresden (IFW Dresden), \\ Helmholtzstraße 20, 01069 Dresden, Germany}

\author{Hrishikesh Borah $^{\dagger\dagger}$}
\affiliation{Max Planck Institute for Chemical Physics of Solids, Nöthnitzer Str. 40, 01187 Dresden, Germany}
\affiliation{Technische Universität Dresden, 01062 Dresden, Germany}

\author{Sandeep Singh}
\affiliation{CSIR - National Physical Laboratory, 
Dr. K.S. Krishnan Marg, 110012 - New Delhi, India}

\author{Berit H. Goodge}
\affiliation{Max Planck Institute for Chemical Physics 
of Solids, Nöthnitzer Str. 40, 01187 Dresden, Germany}

\author{Edouard Lesne}
\affiliation{Max Planck Institute for Chemical Physics 
of Solids, Nöthnitzer Str. 40, 01187 Dresden, Germany}

\author{Sandra Nestler}
\affiliation{Leibniz Institute for Solid State and Materials Research Dresden (IFW Dresden), \\
Helmholtzstraße 20, 01069 Dresden, Germany}

\author{Surinder P. Singh}
\affiliation{CSIR - National Physical Laboratory, 
Dr. K.S. Krishnan Marg, 110012 - New Delhi, India}

\author{Haolin Jin}
\affiliation{Max Planck Institute for Chemical Physics 
of Solids, Nöthnitzer Str. 40, 01187 Dresden, Germany}
\affiliation{Technische Universität Dresden, 01062 Dresden, Germany}

\author{Yejin Lee}
\affiliation{Max Planck Institute for Chemical Physics 
of Solids, Nöthnitzer Str. 40, 01187 Dresden, Germany}

\author{Bernd Büchner}
\affiliation{Leibniz Institute for Solid State and Materials Research Dresden (IFW Dresden), \\ Helmholtzstraße 20, 01069 Dresden, Germany}

\author{Uri Vool $^{\dagger,}$}

\affiliation{Max Planck Institute for Chemical Physics 
of Solids, Nöthnitzer Str. 40, 01187 Dresden, Germany}
\affiliation{Leibniz Institute for Solid State and Materials Research Dresden (IFW Dresden), \\ Helmholtzstraße 20, 01069 Dresden, Germany}

\begin{abstract}
Disorder is a powerful route to high kinetic inductance in superconducting ultrathin films, enabling compact high-impedance quantum circuits. This functionality, however, comes at the cost of reduced phase rigidity and potentially anomalous electrodynamics. Here, we use NbN microwave resonators with thicknesses down to $2.8~\mathrm{nm}$ and sheet kinetic inductance up to $300~\mathrm{pH}/\square$ to probe how this trade-off reshapes the superconducting response. In the thinnest films, transport shows signatures of a Berezinskii-Kosterlitz-Thouless transition, while the microwave response reveals a pronounced low-temperature power-law suppression of the superfluid stiffness, inconsistent with Mattis-Bardeen theory. With increasing thickness, this anomalous regime is progressively suppressed, marking a continuous crossover toward conventional, gap-dominated electrodynamics. Cross-sectional transmission electron microscopy reveals a nanocrystalline twin-domain structure, pointing to oriented microstructural disorder as a crucial factor in the observed response. Overall, the crossover is governed by the ratio of superfluid stiffness to pairing energy, $\Theta(0)/T_c$, identifying this ratio as a parameter governing the boundary between phase-fluctuation-dominated and gap-dominated superconducting electrodynamics in disordered nanofilms.
\end{abstract}
\maketitle
Superconducting quantum circuits rely on microwave components combining low dissipation with large characteristic impedance, compact mode volume, and weak intrinsic nonlinearity~\cite{zmuidzinas2012superconducting, manucharyan2009fluxonium, leduc2010titanium, jouanny2025high,Niepce2019, Frasca2023, masluk2012microwave, bell2012quantum, Wei2023, annunziata2010tunable, khorramshahi2025high, grunhaupt2018loss, grunhaupt2019granular, maleeva2018circuit, rieger2023granular,bottcher2025transmon}. A direct route is to employ inductive elements whose impedance is dominated by kinetic inductance $L_k$~\cite{Frasca2023, Niepce2019, samkharadze2016high, kroll2019magnetic, yu2021magnetic, borisov2020superconducting, roy2026magnetic, frasca2024three, annunziata2010tunable, maleeva2018circuit, winkel2020implementation, yu2024development, ho2012wideband, day2003broadband, parker2022degenerate, vissers2015frequency, adamyan2016tunable, leduc2010titanium, lee1993penetration, khorramshahi2025high, grunhaupt2018loss, rieger2023granular, xu2023magnetic, bottcher2025transmon}, so that the circuit response is set by condensate inertia rather than geometric magnetic energy ~\cite{jouanny2025high,Frasca2023,annunziata2010tunable}. Strongly disordered superconducting thin films are especially attractive in this context: they achieve very large sheet inductance ~\cite{Niepce2019,driessen2012strongly,maleeva2018circuit,grunhaupt2019granular, rieger2023granular,khorramshahi2025high,grunhaupt2018loss,Frasca2023, jouanny2025high, bottcher2025transmon} within a simple, weakly nonlinear material platform ~\cite{Frasca2023,khorramshahi2025high,maleeva2018circuit, jouanny2025high, ho2012wideband}, avoiding the fabrication complexity and residual nonlinearity of Josephson-junction arrays ~\cite{masluk2012microwave,grunhaupt2019granular,bell2012quantum,rieger2023granular,wang2025high}. Among candidate materials, NbN is particularly attractive, combining high $T_c$ and large $L_{k,\square}$ ~\cite{Niepce2019,Frasca2023,Wei2023,xu2023magnetic} with disorder tunable through film thickness ~\cite{chockalingam2008superconducting, sharma2022complex}, allowing it to be driven continuously into the strongly disordered, few-nanometer regime while remaining robustly superconducting \cite{Mondal2011, yong2013robustness, weitzel2023sharpness}.\\
However, the same electronic disorder that enhances $L_k$ also reduces the superfluid stiffness, which is the energy scale governing the rigidity of the superconducting phase. Disorder brings the phase-stiffness scale $\Theta$ closer to the
pairing scale $\Delta$, making phase fluctuations increasingly
important beyond conventional Mattis--Bardeen (M--B) electrodynamics~\cite{Raychaudhuri2022,driessen2012strongly,Eliashberg1991}. In this regime, local Cooper pairing can persist even when global phase coherence is strongly suppressed~\cite{Raychaudhuri2022,Sacepe2010,charpentier2025}. In the ultrathin limit, this separation is realized through the Berezinskii-Kosterlitz-Thouless (BKT) transition, where long-range phase coherence is destroyed by vortex-antivortex unbinding while Cooper pairs remain locally intact~\cite{kosterlitz1973ordering,yong2013robustness,bartolf2010current,benfatto2009broadening,hu2020evidence,Mondal2011,weitzel2023sharpness,sharma2022complex}. Beyond this loss of phase rigidity, spatial variations in the local pairing amplitude can generate low-energy collective modes that are absent in homogeneous superconductors.
These modes provide an additional channel for suppressing the superfluid stiffness at low temperature, causing the microwave response to deviate from the activated behavior expected from thermally excited quasiparticles alone~\cite{driessen2012strongly,Coumou2013,Khvalyuk2024, khvalyuk2026dissipation}. In strongly disordered superconductors, this physics can appear as a power-law suppression of the superfluid stiffness, reflecting spatial variations in the local pairing amplitude~\cite{Khvalyuk2024,Sacepe2010,charpentier2025}.\\
\begin{figure*}[t] 
\begin{minipage}{1.0\textwidth}
\includegraphics[width=\textwidth]{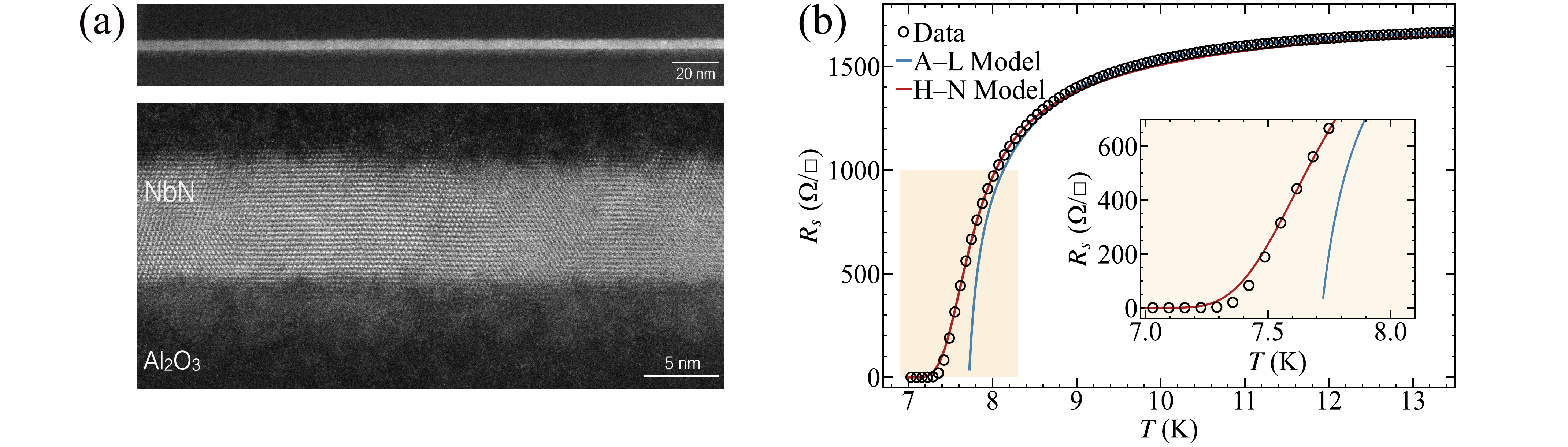}
\end{minipage}
\caption{\justifying (a) Cross-sectional ADF-STEM images of a representative $5~\mathrm{nm}$ NbN film on Al$_2$O$_3$. Top: continuous film over an extended region. Bottom: high-resolution view; the diffuse top layer is the native oxide, while the lattice fringes below confirm crystalline NbN oriented to the substrate. 
(b) Sheet resistance near the superconducting transition of the $2.8~\mathrm{nm}$ film. The red line is an Aslamazov--Larkin fit yielding the mean-field $T_{c}=7.72~\mathrm{K}$; elsewhere, $T_c$ denotes the resistive-transition midpoint. The blue line is a Halperin--Nelson fit yielding $T_{\mathrm{BKT}}=6.959~\mathrm{K}$ (see SM).}
 \label{Figure_1}
\end{figure*}
Ultrathin NbN films let us track this interplay directly, since thickness tunes the disorder while the resonator probes the condensate response. This raises a question that is at once fundamental and technological: can a large kinetic inductance be reached while the condensate remains conventional, or is it inseparable from the anomalous, phase-fluctuation-dominated electrodynamics of disordered systems$?$\\
Here, we realize a thickness-controlled NbN resonator platform spanning film thicknesses from $25$ to $2.8~\mathrm{nm}$ in which the sheet kinetic inductance reaches $300~\mathrm{pH}/\square$, one of the highest values reported for NbN, arising solely from disorder-induced modifications of the film rather than geometric enhancement. The thinnest films exhibit a BKT transition, while the low-temperature fractional resonant frequency shift, $\delta f/f$, departs from Mattis--Bardeen theory, with the anomalous regime narrowing as the film thickness increases. Cross-sectional microscopy points to oriented structural disorder as a key factor, while the ratio $\Theta(0)/T_c$ emerges as the parameter organizing the crossover between phase-fluctuation-dominated and gap-dominated electrodynamics.\\
Structural characterization of the NbN films by X-ray reflectometry revealed a native surface oxide approximately $1.5~\mathrm{nm}$ thick. Cross-sectional scanning transmission electron microscopy (STEM) further showed that, beneath this oxide, the NbN forms a continuous crystalline layer with a well-defined orientation relative to the substrate. [Fig.~\ref{Figure_1}a]. The image shown corresponds to a representative (5~$\mathrm{nm}$) film and is characteristic of the film series. Analysis of Fourier transforms shows that the film reflections are broader than the sharp substrate reflections, and also reveals the coexistence of single-orientation and twin-oriented regions. This identifies the films as substrate-templated nanocrystalline NbN with twin domains, rather than a homogeneously amorphous or randomly polycrystalline solid (see Supplemental Material (SM)). Such structurally correlated disorder, on a scale set by the orientation domains, can give rise to an inhomogeneous pairing landscape — an effect we test directly through the microwave response below.\\
Transport measurements on patterned Hall bars reveal a significant increase in the normal-state sheet resistance, from $128.7~\Omega/\square$ at $d = 25~\mathrm{nm}$ to $1686.4~\Omega/\square$ at $d = 2.8~\mathrm{nm}$~\cite{Kern2024,chockalingam2008superconducting,joshi2018superconducting}. For the thinnest film, whose thickness $d=2.8~\mathrm{nm}$ is smaller than the zero-temperature coherence length $\xi(0)\approx5~\mathrm{nm}$~\cite{semenov2009optical}, the superconducting state is effectively two-dimensional (2-D)~\cite{chu2004phase} and is expected to exhibit a Berezinskii--Kosterlitz--Thouless (BKT) transition~\cite{kosterlitz1973ordering}.
Consistent with this expectation, the $2.8~\mathrm{nm}$ thick film shows a broad superconducting transition [Fig.~\ref{Figure_1}(b)],  which is attributed, in this case, to strong phase fluctuations in the 2-D limit, associated with the unbinding of vortex-antivortex pairs. The excess conductivity above the transition is well described by the Aslamazov--Larkin (AL) form~\cite{aslamasov1968influence} at higher temperatures and crosses over to the Halperin--Nelson (HN) form~\cite{halperin1979resistive} near the transition. This AL-to-HN crossover provides evidence for BKT physics~\cite{venditti2019nonlinear} in the ultrathin film, with the HN fit [Fig.~\ref{Figure_1}(b)] giving $T_{\mathrm{HN}} \equiv T_{\mathrm{BKT}} = 6.959 \pm 0.026~\mathrm{K}$ (for detailed analysis, see SM).\\
For the microwave analysis, we focus on two key scales that set the inductive microwave response: the normal-state sheet resistance $R_\square$, which controls the magnitude of the kinetic inductance, and the mean-field critical temperature $T_{c}$, which sets the pairing gap $\Delta(0) \approx 1.764k_B T_{c}$. The kinetic inductance per square follows from the Mattis--Bardeen relation~\cite{Frasca2023},
\begin{equation}
L_{k,\square}(T)
=\frac{\hbar R_\square}{\pi \Delta(0)}
\frac{1}{\tanh\!\left[\Delta(0)/(2k_B T)\right]},
\label{eq_1}
\end{equation}
which rises at base temperature from $14$ to $300~\mathrm{pH}/\square$ as the thickness is reduced from $25$ to $2.8~\mathrm{nm}$, reflecting the disorder-driven suppression of the superfluid stiffness. Equation~\eqref{eq_1} serves not as a description of the films, but as the Mattis–Bardeen (mean-field BCS) baseline against which the anomalous deviations identified below are measured.\\

\begin{figure*}[t] 
    \centering 
    \begin{minipage}{1.00\textwidth}
        \includegraphics[width=\textwidth]{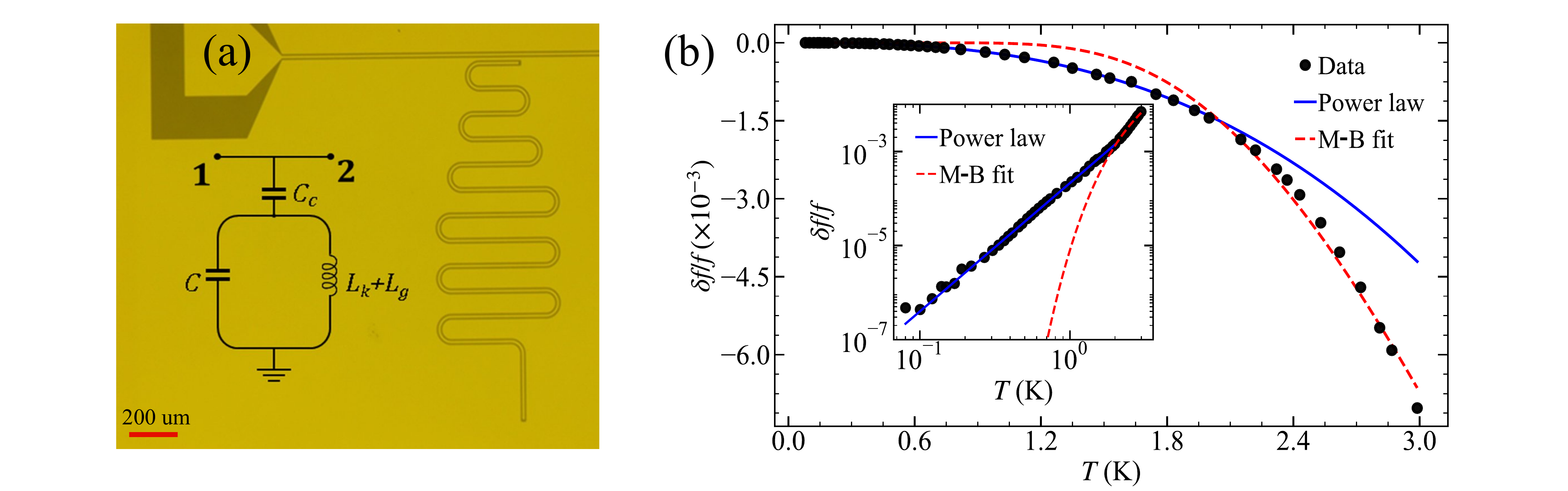}
    \end{minipage}
\caption{\justifying (a) Optical image of the $\lambda/2$ CPW resonator and its lumped-element equivalent circuit.
    (b)~Fractional frequency shift $\delta f/f$ versus temperature on a linear scale, with power-law (blue) and Mattis-Bardeen (red) fits. Inset: log-log representation highlighting the low-temperature scaling.}
    \label{Figure_2}
\end{figure*}
To probe the electrodynamics, each film was patterned into a hanger-type coplanar-waveguide resonator [Fig.~\ref{Figure_2}(a)]. Fits to the complex transmission $S_{21}(f)$, measured near $40~\mathrm{mK}$, yield the resonance frequency $f_0$ and intrinsic quality factor $Q_i$ in the linear, power-independent regime (see SM). The relative contributions of kinetic and geometric inductance are quantified by the kinetic-inductance fraction
$\alpha=L_k/(L_k+L_{\mathrm{geom}})$. Using
$L_{\mathrm{geom}}\approx4.39~\mathrm{pH/\square}$ and the inferred kinetic inductance, we find that $\alpha$ decreases from $0.986$ for the $2.8~\mathrm{nm}$ film to $0.76$ for the $25~\mathrm{nm}$ film. Thus, even the thickest-film resonator remains kinetic-inductance dominated, making the resonance-frequency response directly sensitive to changes in the superfluid stiffness.

We therefore convert the zero-temperature sheet kinetic inductance into the 2-D superfluid stiffness, $\Theta(0)=(\hbar/2e)^2/[k_B L_{k,\square}(0)]$. Having established this absolute stiffness scale, the temperature dependence of the resonance frequency then probes the suppression of this stiffness with increasing temperature. Defining $\delta\Theta(T)=\Theta(0)-\Theta(T)$, the fractional frequency shift is related to the relative change in the 2-D superfluid stiffness as~\cite{Coumou2013,Khvalyuk2024}
\begin{equation}
\frac{\delta f(T)}{f} \approx -\frac{\alpha}{2}\frac{\delta\Theta(T)}{\Theta(0)},
\label{eq:freq_shift}
\end{equation}
where $\alpha$ is the kinetic-inductance participation ratio. Since $\alpha$ is close to unity for the thinnest films, $\delta f/f$ provides a sensitive probe of the condensate electrodynamics.

\begin{figure*}[t] 
    \centering 
    \begin{minipage}{1.0\textwidth}
        \includegraphics[width=\textwidth]{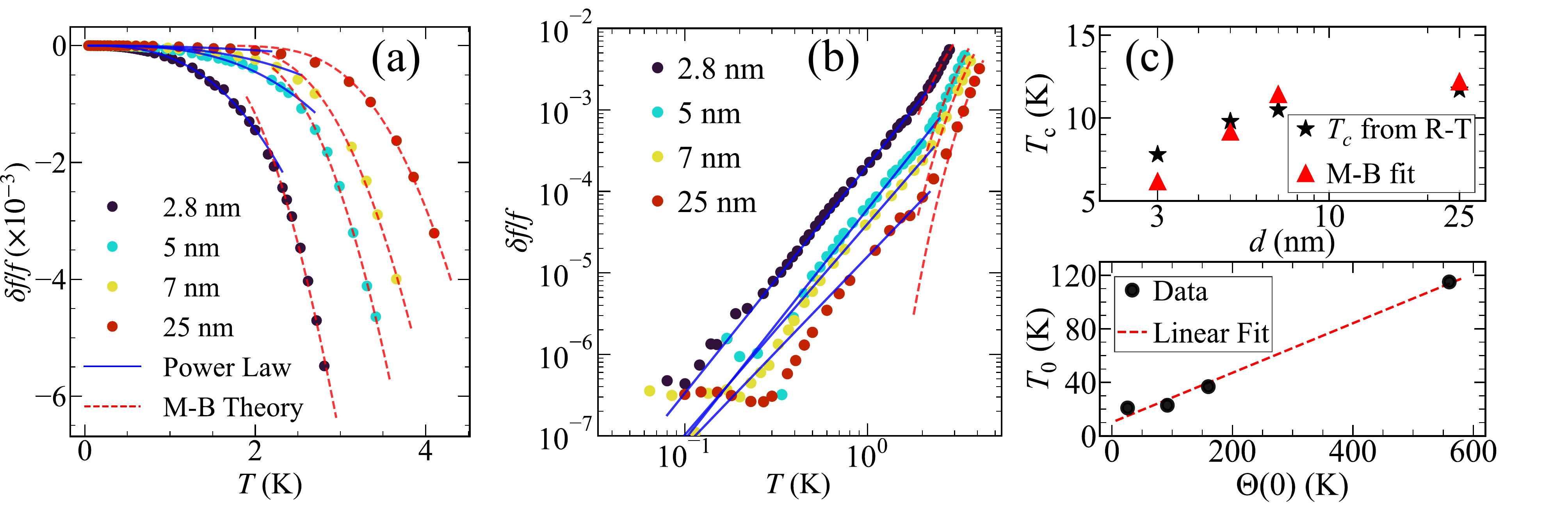}
       \end{minipage}
    \caption{\justifying(a) Fractional resonance frequency shift $\delta f/f$ versus temperature for films with thicknesses ranging from 2.8~nm to 25~nm, with power-law (blue solid lines) and Mattis-Bardeen (red dashed lines) fits. (b) Log-log representation of the same data and fits, highlighting the extended low-temperature power-law scaling. (c) Upper panel: critical temperature $T_c$ extracted from DC transport (black stars) and from the M-B component of the microwave fit (red triangles). Lower panel: Characteristic energy scale $T_0$ as a function of the zero-temperature superfluid stiffness $\Theta(0)$ for films of different thicknesses. Black circles denote values averaged over nominally identical films at each thickness, and the red dashed line is a linear fit.}
    \label{Figure_3} 
\end{figure*}
At low temperatures, $\delta f/f$ for the $2.8~\mathrm{nm}$ resonator deviates strongly from the expected Mattis--Bardeen response and instead follows a robust power-law dependence [Fig.~\ref{Figure_2}(b)],
\begin{equation}
\frac{\delta f}{f} = -\left(\frac{T}{T_0}\right)^b,
\label{eq}
\end{equation}
with $b \approx 2.8$ and $T_0 \approx 20.9~\mathrm{K}$, as evidenced by the extended linear regime in the log--log representation [Fig.~\ref{Figure_2}(b), inset]. Here, $T_0$ is a fit-extracted characteristic scale that enables comparison among thicknesses. Unlike the exponentially activated response expected from thermally excited quasiparticles in M-B electrodynamics, the observed power law points to additional low-energy excitations that remain accessible well below $T_c$. We associate these excitations with collective modes arising from spatial variations in the local pairing amplitude, which can provide an additional channel for depleting the superfluid density. Within this picture, the exponent $b$ reflects the low-energy excitation spectrum: stronger inhomogeneity is expected to enhance the low-energy spectral weight and reduce $b$, whereas a more uniform pairing landscape shifts the response toward the conventional behavior~\cite{Khvalyuk2024}.\\
This interpretation is also consistent with the comparatively large value $b \approx 2.8$ found here, which exceeds those reported for strongly disordered amorphous superconductors~\cite{Khvalyuk2024}. The STEM image in Fig.~\ref{Figure_1}(a) offers a possible explanation. Rather than being dominated by spatially random disorder, NbN forms a crystalline film composed of twin-oriented domains, representing a structurally correlated form of disorder. Such a microstructure may narrow the distribution of local pairing amplitudes and shift the low-energy modes toward higher energies, both of which are consistent with a larger value of $b$. The preservation of a relatively high $T_{c} = 7.72~\mathrm{K}$ at a thickness of only $2.8~\mathrm{nm}$, together with the BKT signatures discussed above, further supports a picture in which local Cooper pairing remains robust even as global phase coherence becomes fragile.\\
At higher temperatures, however, the data progressively depart from the 
power-law form and recover the M-B response as clearly seen in Fig.~\ref{Figure_2}(b).
\[
\left(\frac{\delta f}{f}\right)_{\mathrm{M\text{-}B}} \propto -\sqrt{\frac{\Delta}{k_B T}}\,e^{-\Delta/k_B T},
\]
This crossover indicates a shift toward gap-controlled electrodynamics as quasiparticles increasingly dominate the depletion of the superfluid density. Hence,  the data over the full measured range are well described by the phenomenological crossover form~\cite{Khvalyuk2024},
\begin{equation}
\frac{\delta f}{f} = -\left(\frac{T}{T_0}\right)^b + \left(\frac{\delta f}{f}\right)_{\mathrm{M\text{-}B}},
\label{eq:crossover}
\end{equation}
which captures the smooth evolution from the anomalous low-temperature power law to a conventional M-B behavior at higher temperatures.\\
\begin{table}[t]
\caption{Transport and microwave parameters across the NbN thickness series. $T_0$ and $b$ are extracted from the power-law fit, Eq.~(\ref{eq}).}
\label{tab:params}
\setlength{\tabcolsep}{3pt}
\renewcommand{\arraystretch}{1.15}
\begin{ruledtabular}
\begin{tabular}{cccccccc}
$d$  & $R_{\square}$      & $T_c$ & $L_{k,\square}$ & $\Theta(0)$ & $T_0$ & $b$  & $\frac{\Theta(0)}{T_c}$ \\
(nm) & ($\Omega/\square$) & (K)   & (pH/$\square$)  & (K)         & (K)   &      &                         \\
\hline
2.8  & 1686               & 7.72   & 299             & 26          & 20.9  & 2.79 & 3.4  \\
5    & 600                & 9.8   & 85              & 92          & 22.9  & 2.82 & 9.4  \\
7    & 369                & 10.5  & 49              & 160         & 36.8  & 2.58 & 15   \\
25   & 129                & 11.7  & 14              & 560         & 115   & 2.33 & 48   \\
\end{tabular}
\end{ruledtabular}
\end{table}
Extending this analysis across thickness series, we find that the same two-regime electrodynamic response persists in all films from $2.8$ to $25~\mathrm{nm}$ [Fig.~\ref{Figure_3}(a)]. Each film exhibits a low-temperature power-law dependence followed by a crossover to M-B like behavior at higher temperatures, as further highlighted in log--log representation [Fig.~\ref{Figure_3}(b)]. We find that the value of $b$ remains within a range $\approx 2.3$--$2.8$, with no clear systematic dependence on thickness, despite an order-of-magnitude variation in $R_\square$ and a factor of $\sim 20$ variation in $L_{k,\square}(0)$ (Table~\ref{tab:params}). The exponent $b$ and the characteristic scale $T_0$ therefore appear to play distinct roles: $T_0$ varies strongly with thickness and follows the evolution of the superfluid stiffness $\Theta(0)$, whereas $b$ remains insensitive to the bulk disorder level.\\
An independent consistency check of this two-regime description is provided by the critical temperatures extracted from transport and microwave measurements [Fig.~\ref{Figure_3}(c), upper panel]. The mean-field transition temperature $T_c$ obtained from the higher-temperature M--B component closely agrees with the value extracted from DC transport. In the microwave fit, $T_c$ is determined independently through the gap scale entering the M--B response. Its agreement with transport therefore shows that the higher-temperature contribution is tied to the physical superconducting transition. It is not merely compensating for the low-temperature power-law term. This correspondence supports the decomposition into an anomalous low-temperature power-law regime and a higher-temperature quasiparticle-dominated M--B regime.

Taken together, these results suggest that the shape of the low-temperature anomalous response is not primarily controlled by the overall disorder level, as reflected in $R_\square$ and $L_{k,\square}(0)$. Instead, the dominant thickness-dependent change appears in the stiffness-related scale $T_0$. This separation between a nearly unchanged exponent and a strongly varying stiffness scale is consistent with a scenario in which the low-energy excitation spectrum is influenced by a common microstructural motif across the series, namely the substrate-templated nanocrystalline layer with twin-oriented domains identified in Fig.~\ref{Figure_1}(a). In this sense, NbN realizes a regime in which the phase-stiffness scale can be tuned substantially with thickness, from the strongly phase-fluctuating ultrathin limit toward a more gap-dominated $25~\mathrm{nm}$ film, while the form of the low-temperature power-law response remains comparatively robust. This contrasts with homogeneously disordered systems, where the exponent $b$ is expected to vary more directly with disorder strength \cite{Khvalyuk2024}.
This crossover is also reflected in the temperature range over which the anomalous response persists. The power-law regime is observed up to $T \lesssim 0.35 T_c$ in the thinnest $2.8~\mathrm{nm}$ film, whereas in the thickest $25~\mathrm{nm}$ film it is restricted to $T \lesssim 0.15 T_c$ as clearly observed in Fig.~\ref{Figure_3}(b). This trend is consistent with the picture that reduced superfluid stiffness extends the temperature window over which phase fluctuations dominate. Since $L_{k,\square}(0)$ is inversely related to the superfluid stiffness, the same trend appears as an anti-correlation between $T_0$ and $L_{k,\square}(0)$ (see Table~\ref{tab:params}).\\
As shown in [Fig.~\ref{Figure_3}(c), lower panel], $T_0$ increases monotonically with the zero-temperature superfluid stiffness $\Theta(0)$, consistent with a phase-fluctuation contribution to the anomalous response. The same trend is reproduced across several nominally identical films measured at each thickness. In strongly disordered superconductors, $T_0$ has been found to obey an approximately stiffness-governed proportionality, $T_0\propto\Theta(0)$, suggesting that a single stiffness scale can control the anomalous electrodynamic response~\cite{Khvalyuk2024}.
In NbN, however, the dependence on stiffness is weaker and appears to involve an additional energy scale. Across our thickness series, $\Theta(0)$ varies by more than an order of magnitude, whereas $T_c$ changes by less than a factor of two. Consequently, $\Theta(0)/T_c$ spans from $3.4$ to $48$ (Table~\ref{tab:params}), allowing the stiffness and pairing scales to be varied partially independently. Over this range, a stiffness-only proportionality does not fully capture the data. Instead, the correlation is better described empirically by the linear relation
\begin{equation}
T_0 \approx 0.18\Theta(0)+T_{\mathrm{off}},
\qquad
T_{\mathrm{off}}\approx10~\mathrm{K}.
\label{eq:T0fit}
\end{equation}
The finite offset is suggestive because $T_{\mathrm{off}}$ is comparable to the average critical temperature of the NbN films. This raises the possibility that it reflects a pairing-related energy scale, set by $\Delta\propto k_B T_c$, rather than merely an arbitrary fitting constant.
Within this interpretation, $T_0$ is not explicitly controlled by the superfluid stiffness. Instead, the anomalous electrodynamic scale may depend on both the phase stiffness $\Theta(0)$ and the pairing scale $\Delta$. A stiffness-only description may therefore be sufficient in strongly disordered systems where these scales nearly track each other, whereas NbN lies in an intermediate regime where they are partially decoupled. With this caveat, we identify $\Theta(0)/T_c$ as a useful parameter for classifying the crossover between phase-fluctuation-dominated electrodynamics in the thinnest films and more gap-dominated behavior in the thicker films. A broader disorder range and larger sample set will be needed to establish the microscopic origin of the offset more definitively.\\
Microwave-loss measurements, discussed in the End Matter, further show that TLS dissipation is strongest in the extreme ultrathin limit, whereas thicker films retain substantially higher $Q_i$ and weaker power dependence, identifying a regime in which large kinetic inductance can coexist with comparatively low microwave loss.\\
In summary, we have fabricated and characterized microwave resonators from NbN films, a material well suited to high-impedance circuit elements. Even into the few-nanometer regime the devices 
retain clear, well-defined microwave behavior, showing that large kinetic 
inductance can be obtained intrinsically from the material while preserving 
useful device performance, and the series let us track the sheet kinetic 
inductance and superfluid stiffness with thickness.\\
These same resonators also probe the superconducting condensate itself. The temperature 
dependence of the resonance frequency reveals a robust low-temperature power law 
that conventional Mattis--Bardeen electrodynamics cannot account for alone, 
placing the films in an intermediate-disorder regime where the phase-stiffness 
scale $\Theta(0)$ and the pairing scale $\Delta$ are no longer well separated and 
both may contribute. The crossover from phase-fluctuation-dominated behavior in 
the thinnest films to gap-dominated behavior in the thicker ones is demarcated by 
the single ratio $\Theta(0)/T_c$. We expect this work to be useful both for 
scalable high-impedance circuit elements and for fundamental efforts to 
understand how pairing, phase stiffness, and collective modes shape the 
electrodynamics of superconducting ultrathin films.\\
\begin{acknowledgments}
\noindent
We thank Andrea Perali, Christopher Strunk, Nicola Poccia, and Ioan Pop for stimulating discussions. This work was funded by the European Union through the ERC Starting Grant cQEDscope (Grant Agreement No. 101075962) and was partially supported by the Deutsche Forschungsgemeinschaft (DFG, German Research Foundation) under Project No. 539383397. Additional support was provided through the GAP180932 and MLP190932 projects of the Council of Scientific and Industrial Research (CSIR), India, under the Ministry of Commerce and Industry.

\end{acknowledgments}


\bibliographystyle{apsrev4-2}  
\bibliography{Power_Law.bib}  

@article{khorramshahi2025high,
  title={High-impedance granular-aluminum ring resonators},
  author={Khorramshahi, Mahya and Spiecker, Martin and Paluch, Patrick and Geisert, Simon and Gosling, Nicolas and Zapata, Nicolas and Brauch, Lucas and K{\"u}bel, Christian and Dehm, Simone and Krupke, Ralph and others},
  journal={Physical Review Applied},
  volume={24},
  number={2},
  pages={024066},
  year={2025},
  publisher={APS}
}

@article{jouanny2025high,
  title={High kinetic inductance cavity arrays for compact band engineering and topology-based disorder meters},
  author={Jouanny, Vincent and Frasca, Simone and Weibel, Vera Jo and Peyruchat, L{\'e}o and Scigliuzzo, Marco and Oppliger, Fabian and De Palma, Franco and Sbroggi{\`o}, Davide and Beaulieu, Guillaume and Zilberberg, Oded and others},
  journal={Nature Communications},
  volume={16},
  number={1},
  pages={3396},
  year={2025},
  publisher={Nature Publishing Group UK London}
}

@article{driessen2012strongly,
  title={Strongly Disordered TiN and NbTiN s-Wave Superconductors<? format?> Probed by Microwave Electrodynamics},
  author={Driessen, Eduard FC and Coumou, PCJJ and Tromp, RR and De Visser, PJ and Klapwijk, TM},
  journal={Physical Review Letters},
  volume={109},
  number={10},
  pages={107003},
  year={2012},
  publisher={APS}
}

@article{Niepce2019,
  author = {Niepce, David and Burnett, Jonathan and Bylander, Jonas},
  title = {High Kinetic Inductance NbN Nanowire Superinductors},
  journal = {Physical Review Applied},
  volume = {11},
  pages = {044014},
  year = {2019},
  doi = {10.1103/PhysRevApplied.11.044014}
}

@article{Frasca2023,
  author = {Frasca, S. and Arabadzhiev, I. N. and Bros de Puechredon, S. Y. and Oppliger, F. and Jouanny, V. and Musio, R. and Scigliuzzo, M. and Minganti, F. and Scarlino, P. and Charbon, E.},
  title = {NbN films with high kinetic inductance for high-quality compact superconducting resonators},
  journal = {Physical Review Applied},
  volume = {20},
  pages = {044021},
  year = {2023},
  doi = {10.1103/PhysRevApplied.20.044021}
}

@article{masluk2012microwave,
  title={Microwave characterization of josephson junction arrays: Implementing<? format?> a low loss superinductance},
  author={Masluk, Nicholas A and Pop, Ioan M and Kamal, Archana and Minev, Zlatko K and Devoret, Michel H},
  journal={Physical Review Letters},
  volume={109},
  number={13},
  pages={137002},
  year={2012},
  publisher={APS}
}

@article{bell2012quantum,
  title={Quantum superinductor with tunable nonlinearity},
  author={Bell, MT and Sadovskyy, IA and Ioffe, LB and Kitaev, A Yu and Gershenson, ME},
  journal={Physical Review Letters},
  volume={109},
  number={13},
  pages={137003},
  year={2012},
  publisher={APS}
}

@article{winkel2020implementation,
  title={Implementation of a transmon qubit using superconducting granular aluminum},
  author={Winkel, Patrick and Borisov, Kiril and Gr{\"u}nhaupt, Lukas and Rieger, Dennis and Spiecker, Martin and Valenti, Francesco and Ustinov, Alexey V and Wernsdorfer, Wolfgang and Pop, Ioan M},
  journal={Physical Review X},
  volume={10},
  number={3},
  pages={031032},
  year={2020},
  publisher={APS}
}

@article{day2003broadband,
  title={A broadband superconducting detector suitable for use in large arrays},
  author={Day, Peter K and LeDuc, Henry G and Mazin, Benjamin A and Vayonakis, Anastasios and Zmuidzinas, Jonas},
  journal={Nature},
  volume={425},
  number={6960},
  pages={817--821},
  year={2003},
  publisher={Nature Publishing Group UK London}
}

@article{ho2012wideband,
  title={A wideband, low-noise superconducting amplifier with high dynamic range},
  author={Ho Eom, Byeong and Day, Peter K and LeDuc, Henry G and Zmuidzinas, Jonas},
  journal={Nature Physics},
  volume={8},
  number={8},
  pages={623--627},
  year={2012},
  publisher={Nature Publishing Group UK London}
}

@article{Wei2023,
  author = {Wei, XingYu and Jiang, JunLiang and Xu, Wenqu and Guo, Tingting and Zhang, Kaixuan and Li, Zishuo and Zhou, Tianshi and Sheng, Yifan and Cao, Chunhai and Sun, Guozhu and Wu, Peiheng},
  title = {Compact superconducting transmon qubit circuits made of ultrathin NbN},
  journal = {Applied Physics Letters},
  volume = {123},
  pages = {154005},
  year = {2023},
  doi = {10.1063/5.0170259}
}

@article{Raychaudhuri2022,
  author = {Raychaudhuri, Pratap and Dutta, Surajit},
  title = {Phase fluctuations in conventional superconductors},
  journal = {Journal of Physics: Condensed Matter},
  volume = {34},
  pages = {083001},
  year = {2022},
  doi = {10.1088/1361-648X/ac360b}
}

@article{Eliashberg1991,
  author = {Eliashberg, G. M. and Klimovitch, G. V. and Rylyakov, A. V.},
  title = {On the Temperature Dependence of the London Penetration Depth in a Superconductor},
  journal = {Journal of Superconductivity},
  volume = {4},
  pages = {393--396},
  year = {1991},
  doi = {10.1007/BF00618337}
}

@article{Sacepe2010,
  author = {Sac{\'e}p{\'e}, B. and Chapelier, C. and Baturina, T. I. and Vinokur, V. M. and Baklanov, M. R. and Sanquer, M.},
  title = {Pseudogap in a thin film of a conventional superconductor},
  journal = {Nature Communications},
  volume = {1},
  pages = {140},
  year = {2010},
  doi = {10.1038/ncomms1132}
}

@article{Coumou2013,
  author = {Coumou, P. C. J. J. and Baryshev, A. M. and Golubov, A. A. and Rogalla, H. and Klapwijk, T. M.},
  title = {Electrodynamic response and local tunneling spectroscopy of strongly disordered superconducting TiN films},
  journal = {Physical Review B},
  volume = {88},
  pages = {180505},
  year = {2013},
  doi = {10.1103/PhysRevB.88.180505}
}

@article{Khvalyuk2024,
  author = {Khvalyuk, Anton V. and Charpentier, Thibault and Roch, Nicolas and Sac{\'e}p{\'e}, Benjamin and Feigel'man, Mikhail V.},
  title = {Near power-law temperature dependence of the superfluid stiffness in strongly disordered superconductors},
  journal = {Physical Review B},
  volume = {109},
  pages = {144501},
  year = {2024},
  doi = {10.1103/PhysRevB.109.144501}
}

@article{Mondal2011,
  author = {Mondal, M. and Kamlapure, A. and Chand, M. and Saraswat, G. and Kumar, S. and Jesudasan, J. and Benfatto, L. and Tripathi, V. and Raychaudhuri, P.},
  title = {Phase fluctuations in a strongly disordered s-wave NbN superconductor close to the metal-insulator transition},
  journal = {Physical Review Letters},
  volume = {106},
  pages = {047001},
  year = {2011},
  doi = {10.1103/PhysRevLett.106.047001}
}

@article{halperin1979resistive,
  title={Resistive transition in superconducting films},
  author={Halperin, BI and Nelson, David R},
  journal={Journal of low temperature physics},
  volume={36},
  number={5},
  pages={599--616},
  year={1979},
  publisher={Springer}
}

@article{Kern2024,
  author = {Kern, S. and Neilinger, P. and Pol{\'a}{\v c}kov{\'a}, M. and Bar{\'a}nek, T. and Plecenik, T. and Roch, T. and Grajcar, M.},
  title = {Optical and transport properties of NbN thin films revisited},
  journal = {Physical Review B},
  volume = {110},
  pages = {245131},
  year = {2024},
  doi = {10.1103/PhysRevB.110.245131}
}

@article{kosterlitz1973ordering,
  title={Ordering, metastability and phase transitions in two-dimensional systems},
  author={Kosterlitz, John Michael and Thouless, David James},
  journal={Journal of Physics C: Solid State Physics},
  volume={6},
  number={7},
  pages={1181--1203},
  year={1973}
}

@article{yong2013robustness,
  title={Robustness of the Berezinskii-Kosterlitz-Thouless transition in ultrathin NbN films near the superconductor-insulator transition},
  author={Yong, Jie and Lemberger, TR and Benfatto, L and Ilin, K and Siegel, M},
  journal={Physical Review B—Condensed Matter and Materials Physics},
  volume={87},
  number={18},
  pages={184505},
  year={2013},
  publisher={APS}
}

@article{bartolf2010current,
  title={Current-assisted thermally activated flux liberation in ultrathin nanopatterned NbN superconducting meander structures},
  author={Bartolf, Holger and Engel, Andreas and Schilling, Andreas and Il’in, Konstantin and Siegel, Michael and H{\"u}bers, H-W and Semenov, A},
  journal={Physical Review B—Condensed Matter and Materials Physics},
  volume={81},
  number={2},
  pages={024502},
  year={2010},
  publisher={APS}
}

@article{hu2020evidence,
  title={Evidence of the Berezinskii-Kosterlitz-Thouless phase in a frustrated magnet},
  author={Hu, Ze and Ma, Zhen and Liao, Yuan-Da and Li, Han and Ma, Chunsheng and Cui, Yi and Shangguan, Yanyan and Huang, Zhentao and Qi, Yang and Li, Wei and others},
  journal={Nature communications},
  volume={11},
  number={1},
  pages={5631},
  year={2020},
  publisher={Nature Publishing Group UK London}
}

@article{benfatto2009broadening,
  title={Broadening of the Berezinskii-Kosterlitz-Thouless superconducting transition by inhomogeneity and finite-size effects},
  author={Benfatto, L and Castellani, Claudio and Giamarchi, Thierry},
  journal={Physical Review B—Condensed Matter and Materials Physics},
  volume={80},
  number={21},
  pages={214506},
  year={2009},
  publisher={APS}
}

@article{weitzel2023sharpness,
  title={Sharpness of the Berezinskii-Kosterlitz-Thouless transition in disordered NbN films},
  author={Weitzel, Alexander and Pfaffinger, Lea and Maccari, Ilaria and Kronfeldner, Klaus and Huber, Thomas and Fuchs, Lorenz and Mallord, James and Linzen, Sven and Il’ichev, Evgeni and Paradiso, Nicola and others},
  journal={Physical Review Letters},
  volume={131},
  number={18},
  pages={186002},
  year={2023},
  publisher={APS}
}

@article{sharma2022complex,
  title={Complex phase-fluctuation effects correlated with granularity in superconducting NbN nanofilms},
  author={Sharma, Meenakshi and Singh, Manju and Rakshit, Rajib K and Singh, Surinder P and Fretto, Matteo and De Leo, Natascia and Perali, Andrea and Pinto, Nicola},
  journal={Nanomaterials},
  volume={12},
  number={23},
  pages={4109},
  year={2022},
  publisher={MDPI}
}

@article{chockalingam2008superconducting,
  title={Superconducting properties and Hall effect of epitaxial NbN thin films},
  author={Chockalingam, SP and Chand, Madhavi and Jesudasan, John and Tripathi, Vikram and Raychaudhuri, Pratap},
  journal={Physical Review B—Condensed Matter and Materials Physics},
  volume={77},
  number={21},
  pages={214503},
  year={2008},
  publisher={APS}
}

@article{joshi2018superconducting,
  title={Superconducting properties of NbN film, bridge and meanders},
  author={Joshi, Lalit M and Verma, Apoorva and Gupta, Anurag and Rout, PK and Husale, Sudhir and Budhani, RC},
  journal={AIP Advances},
  volume={8},
  number={5},
  year={2018},
  publisher={AIP Publishing}
}

@article{chu2004phase,
  title={Phase slips in superconducting films with constrictions},
  author={Chu, Sang L and Bollinger, Anthony Travis and Bezryadin, Alexey},
  journal={Physical Review B—Condensed Matter and Materials Physics},
  volume={70},
  number={21},
  pages={214506},
  year={2004},
  publisher={APS}
}

@article{aslamasov1968influence,
  title={The influence of fluctuation pairing of electrons on the conductivity of normal metal},
  author={Aslamasov, LG and Larkin, AI},
  journal={Physics Letters A},
  volume={26},
  number={6},
  pages={238--239},
  year={1968},
  publisher={Elsevier}
}

@article{nelson1977universal,
  title={Universal jump in the superfluid density of two-dimensional superfluids},
  author={Nelson, David R and Kosterlitz, John Michael},
  journal={Physical Review Letters},
  volume={39},
  number={19},
  pages={1201},
  year={1977},
  publisher={APS}
}

@article{sharma2026berezinskii,
  title={Berezinskii-Kosterlitz-Thouless to BCS-like superconducting fluctuation crossover driven by weak magnetic fields in ultra-thin NbN films},
  author={Sharma, Meenakshi and Caprara, Sergio and Perali, Andrea and Singh, Surinder P and Singh, Sandeep and Fretto, Matteo and De Leo, Natascia and Pinto, Nicola},
  journal={Superconductivity},
  pages={100256},
  year={2026},
  publisher={Elsevier}
}

@article{zmuidzinas2012superconducting,
  title={Superconducting microresonators: Physics and applications},
  author={Zmuidzinas, Jonas},
  journal={Annu. Rev. Condens. Matter Phys.},
  volume={3},
  number={1},
  pages={169--214},
  year={2012},
  publisher={Annual Reviews}
}

@article{manucharyan2009fluxonium,
  title={Fluxonium: Single cooper-pair circuit free of charge offsets},
  author={Manucharyan, Vladimir E and Koch, Jens and Glazman, Leonid I and Devoret, Michel H},
  journal={Science},
  volume={326},
  number={5949},
  pages={113--116},
  year={2009},
  publisher={American Association for the Advancement of Science}
}

@article{grunhaupt2018loss,
  title={Loss mechanisms and quasiparticle dynamics in superconducting microwave resonators made of thin-film granular aluminum},
  author={Gr{\"u}nhaupt, Lukas and Maleeva, Nataliya and Skacel, Sebastian T and Calvo, Martino and Levy-Bertrand, Florence and Ustinov, Alexey V and Rotzinger, Hannes and Monfardini, Alessandro and Catelani, Gianluigi and Pop, Ioan M},
  journal={Physical Review Letters},
  volume={121},
  number={11},
  pages={117001},
  year={2018},
  publisher={APS}
}

@article{leduc2010titanium,
  title={Titanium nitride films for ultrasensitive microresonator detectors},
  author={Leduc, Henry G and Bumble, Bruce and Day, Peter K and Eom, Byeong Ho and Gao, Jiansong and Golwala, Sunil and Mazin, Benjamin A and McHugh, Sean and Merrill, Andrew and Moore, David C and others},
  journal={Applied Physics Letters},
  volume={97},
  number={10},
  year={2010},
  publisher={AIP Publishing}
}

@article{annunziata2010tunable,
  title={Tunable superconducting nanoinductors},
  author={Annunziata, Anthony J and Santavicca, Daniel F and Frunzio, Luigi and Catelani, Gianluigi and Rooks, Michael J and Frydman, Aviad and Prober, Daniel E},
  journal={Nanotechnology},
  volume={21},
  number={44},
  pages={445202},
  year={2010}
}

@article{maleeva2018circuit,
  title={Circuit quantum electrodynamics of granular aluminum resonators},
  author={Maleeva, Nataliya and Gr{\"u}nhaupt, Lukas and Klein, T and Levy-Bertrand, F and Dupre, O and Calvo, M and Valenti, F and Winkel, P and Friedrich, F and Wernsdorfer, W and others},
  journal={Nature communications},
  volume={9},
  number={1},
  pages={3889},
  year={2018},
  publisher={Nature Publishing Group UK London}
}

@article{rieger2023granular,
  title={Granular aluminium nanojunction fluxonium qubit},
  author={Rieger, D and G{\"u}nzler, S and Spiecker, M and Paluch, P and Winkel, P and Hahn, L and Hohmann, JK and Bacher, A and Wernsdorfer, W and Pop, IM},
  journal={Nature Materials},
  volume={22},
  number={2},
  pages={194--199},
  year={2023},
  publisher={Nature Publishing Group UK London}
}

@article{bottcher2025transmon,
  title={A transmon qubit realized by exploiting the superconductor-insulator transition},
  author={B{\o}ttcher, CGL and {\"O}nder, E and Connolly, T and Zhao, J and Kvande, C and Wang, DQ and Kurilovich, PD and Vaitiek{\.e}nas, S and Glazman, LI and Tang, HX and others},
  journal={arXiv preprint arXiv:2510.19983},
  year={2025}
}

@article{samkharadze2016high,
  title={High-kinetic-inductance superconducting nanowire resonators for circuit QED in a magnetic field},
  author={Samkharadze, Nodar and Bruno, A and Scarlino, Pasquale and Zheng, G and DiVincenzo, DP and DiCarlo, L and Vandersypen, LMK},
  journal={Physical Review Applied},
  volume={5},
  number={4},
  pages={044004},
  year={2016},
  publisher={APS}
}

@article{kroll2019magnetic,
  title={Magnetic-field-resilient superconducting coplanar-waveguide resonators for hybrid circuit quantum electrodynamics experiments},
  author={Kroll, James G and Borsoi, F and Van Der Enden, KL and Uilhoorn, W and De Jong, D and Quintero-P{\'e}rez, M and Van Woerkom, DJ and Bruno, A and Plissard, SR and Car, D and others},
  journal={Physical Review Applied},
  volume={11},
  number={6},
  pages={064053},
  year={2019},
  publisher={APS}
}

@article{yu2021magnetic,
  title={Magnetic field resilient high kinetic inductance superconducting niobium nitride coplanar waveguide resonators},
  author={Yu, C{\'e}cile Xinqing and Zihlmann, Simon and Troncoso Fern{\'a}ndez-Bada, Gonzalo and Thomassin, Jean-Luc and Gustavo, Fr{\'e}d{\'e}ric and Dumur, {\'E}tienne and Maurand, Romain},
  journal={Applied Physics Letters},
  volume={118},
  number={5},
  year={2021},
  publisher={AIP Publishing}
}

@article{borisov2020superconducting,
  title={Superconducting granular aluminum resonators resilient to magnetic fields up to 1 Tesla},
  author={Borisov, K and Rieger, D and Winkel, P and Henriques, F and Valenti, F and Ionita, A and Wessbecher, M and Spiecker, M and Gusenkova, D and Pop, IM and others},
  journal={Applied Physics Letters},
  volume={117},
  number={12},
  year={2020},
  publisher={AIP Publishing}
}

@article{roy2026magnetic,
  title={Magnetic-field-resilient high-impedance high-kinetic-inductance superconducting resonators},
  author={Roy, Camille and Frasca, Simone and Scarlino, Pasquale},
  journal={Physical Review Applied},
  volume={25},
  number={1},
  pages={014069},
  year={2026},
  publisher={APS}
}

@article{frasca2024three,
  title={Three-wave-mixing quantum-limited kinetic inductance parametric amplifier operating at 6 T near 1 K},
  author={Frasca, Simone and Roy, Camille and Beaulieu, Guillaume and Scarlino, Pasquale},
  journal={Physical Review Applied},
  volume={21},
  number={2},
  pages={024011},
  year={2024},
  publisher={APS}
}

@article{yu2024development,
  title={Development of kinetic inductance bolometers for the space high-cadence observing telescope},
  author={Yu, Shiling and Yang, Lihui and Yan, Xiaohui and Fan, Ruirui and Dai, Xucheng and Mai, Zhanzhang and Shi, Zhongyu and Wang, Yu and Zhang, Mingzhu and Hong, Yue and others},
  journal={Journal of Low Temperature Physics},
  volume={215},
  number={3},
  pages={276--284},
  year={2024},
  publisher={Springer}
}

@article{parker2022degenerate,
  title={Degenerate parametric amplification via three-wave mixing using kinetic inductance},
  author={Parker, Daniel J and Savytskyi, Mykhailo and Vine, Wyatt and Laucht, Arne and Duty, Timothy and Morello, Andrea and Grimsmo, Arne L and Pla, Jarryd J},
  journal={Physical Review Applied},
  volume={17},
  number={3},
  pages={034064},
  year={2022},
  publisher={APS}
}

@article{vissers2015frequency,
  title={Frequency-tunable superconducting resonators via nonlinear kinetic inductance},
  author={Vissers, Michael R and Hubmayr, Johannes and Sandberg, Martin and Chaudhuri, Saptarshi and Bockstiegel, Clint and Gao, Jiansong},
  journal={Applied Physics Letters},
  volume={107},
  number={6},
  year={2015},
  publisher={AIP Publishing}
}

@article{adamyan2016tunable,
  title={Tunable superconducting microstrip resonators},
  author={Adamyan, AA and Kubatkin, SE and Danilov, AV},
  journal={Applied Physics Letters},
  volume={108},
  number={17},
  year={2016},
  publisher={AIP Publishing}
}

@article{lee1993penetration,
  title={Penetration depth $\lambda$ (T) of YBa2Cu3O7- $\delta$ films determined from the kinetic inductance},
  author={Lee, JuYoung and Lemberger, Thomas R},
  journal={Applied physics letters},
  volume={62},
  number={19},
  pages={2419--2421},
  year={1993},
  publisher={American Institute of Physics}
}

@article{grunhaupt2019granular,
  title={Granular aluminium as a superconducting material for high-impedance quantum circuits},
  author={Gr{\"u}nhaupt, Lukas and Spiecker, Martin and Gusenkova, Daria and Maleeva, Nataliya and Skacel, Sebastian T and Takmakov, Ivan and Valenti, Francesco and Winkel, Patrick and Rotzinger, Hannes and Wernsdorfer, Wolfgang and others},
  journal={Nature materials},
  volume={18},
  number={8},
  pages={816--819},
  year={2019},
  publisher={Nature Publishing Group UK London}
}

@article{charpentier2025,
  title={First-order quantum breakdown of superconductivity in an amorphous superconductor},
  author={Charpentier, Thibault and Perconte, David and L{\'e}ger, S{\'e}bastien and Amin, Kazi Rafsanjani and Blondelle, Florent and Gay, Fr{\'e}d{\'e}ric and Buisson, Olivier and Ioffe, Lev and Khvalyuk, Anton and Poboiko, Igor and others},
  journal={Nature Physics},
  volume={21},
  number={1},
  pages={104--109},
  year={2025},
  publisher={Nature Publishing Group UK London}
}

@article{wang2025high,
  title={High-coherence fluxonium qubits manufactured with a wafer-scale-uniformity process},
  author={Wang, Fei and Lu, Kannan and Zhan, Huijuan and Ma, Lu and Wu, Feng and Sun, Hantao and Deng, Hao and Bai, Yang and Bao, Feng and Chang, Xu and others},
  journal={Physical Review Applied},
  volume={23},
  number={4},
  pages={044064},
  year={2025},
  publisher={APS}
}

@article{xu2023magnetic,
  title={Magnetic field-resilient quantum-limited parametric amplifier},
  author={Xu, Mingrui and Cheng, Risheng and Wu, Yufeng and Liu, Gangqiang and Tang, Hong X},
  journal={PRX Quantum},
  volume={4},
  number={1},
  pages={010322},
  year={2023},
  publisher={APS}
}

@article{muller2019towards,
  title={Towards understanding two-level-systems in amorphous solids: insights from quantum circuits},
  author={M{\"u}ller, Clemens and Cole, Jared H and Lisenfeld, J{\"u}rgen},
  journal={Reports on Progress in Physics},
  volume={82},
  number={12},
  pages={124501},
  year={2019},
  publisher={IOP Publishing}
}

@article{megrant2012planar,
  title={Planar superconducting resonators with internal quality factors above one million},
  author={Megrant, Anthony and Neill, Charles and Barends, Rami and Chiaro, Ben and Chen, Yu and Feigl, Ludwig and Kelly, Julian and Lucero, Erik and Mariantoni, Matteo and O’Malley, Peter JJ and others},
  journal={Applied Physics Letters},
  volume={100},
  number={11},
  year={2012},
  publisher={AIP Publishing}
}

@article{mcrae2020materials,
  title={Materials loss measurements using superconducting microwave resonators},
  author={McRae, Corey Rae Harrington and Wang, Haozhi and Gao, Jiansong and Vissers, Michael R and Brecht, Teresa and Dunsworth, Andrew and Pappas, David P and Mutus, Josh},
  journal={Review of Scientific Instruments},
  volume={91},
  number={9},
  year={2020},
  publisher={AIP Publishing}
}

@article{pappas2011two,
  title={Two level system loss in superconducting microwave resonators},
  author={Pappas, David P and Vissers, Michael R and Wisbey, David S and Kline, Jeffrey S and Gao, Jiansong},
  journal={IEEE Transactions on Applied Superconductivity},
  volume={21},
  number={3},
  pages={871--874},
  year={2011},
  publisher={IEEE}
}

@article{kumar2008temperature,
  title={Temperature dependence of the frequency and noise of superconducting coplanar waveguide resonators},
  author={Kumar, Shwetank and Gao, Jiansong and Zmuidzinas, Jonas and Mazin, Benjamin A and LeDuc, Henry G and Day, Peter K},
  journal={Applied Physics Letters},
  volume={92},
  number={12},
  year={2008},
  publisher={AIP Publishing}
}

@article{kirsh2017revealing,
  title={Revealing the nonlinear response of a tunneling two-level system ensemble using coupled modes},
  author={Kirsh, Naftali and Svetitsky, Elisha and Burin, Alexander L and Schechter, Moshe and Katz, Nadav},
  journal={Physical Review Materials},
  volume={1},
  number={1},
  pages={012601},
  year={2017},
  publisher={APS}
}

@article{semenov2009optical,
  title={Optical and transport properties of ultrathin NbN films and nanostructures},
  author={Semenov, Alexej and G{\"u}nther, Burghardt and B{\"o}ttger, Ute and H{\"u}bers, H-W and Bartolf, Holger and Engel, Andreas and Schilling, Andreas and Ilin, Konstantin and Siegel, Michael and Schneider, R and others},
  journal={Physical Review B—Condensed Matter and Materials Physics},
  volume={80},
  number={5},
  pages={054510},
  year={2009},
  publisher={APS}
}

@article{venditti2019nonlinear,
  title={Nonlinear I-V characteristics of two-dimensional superconductors: Berezinskii-Kosterlitz-Thouless physics versus inhomogeneity},
  author={Venditti, G and Biscaras, J and Hurand, S and Bergeal, N and Lesueur, J and Dogra, A and Budhani, RC and Mondal, Mintu and Jesudasan, John and Raychaudhuri, Pratap and others},
  journal={Physical Review B},
  volume={100},
  number={6},
  pages={064506},
  year={2019},
  publisher={APS}
}

@article{khvalyuk2026dissipation,
  title={Dissipation due to bulk localized low-energy modes in strongly disordered superconductors},
  author={Khvalyuk, Anton V and Feigel’man, Mikhail V},
  journal={Physical Review Letters},
  volume={136},
  number={25},
  pages={256001},
  year={2026},
  publisher={APS}
}

@article{martinis2005decoherence,
  title={Decoherence in Josephson qubits from dielectric loss},
  author={Martinis, John M and Cooper, Ken B and McDermott, Robert and Steffen, Matthias and Ansmann, Markus and Osborn, KD and Cicak, Katarina and Oh, Seongshik and Pappas, David P and Simmonds, Raymond W and others},
  journal={Physical Review Letters},
  volume={95},
  number={21},
  pages={210503},
  year={2005},
  publisher={APS}
}

@article{mattis1958theory,
  title={Theory of the anomalous skin effect in normal and superconducting metals},
  author={Mattis, Daniel C and Bardeen, John},
  journal={Physical Review},
  volume={111},
  number={2},
  pages={412},
  year={1958},
  publisher={APS}
}
\begin{center}
\textbf{End Matter: Microwave Loss Mechanisms}
\end{center}
\begin{figure*}[t]
    \centering
    \includegraphics[width=1.00\textwidth]{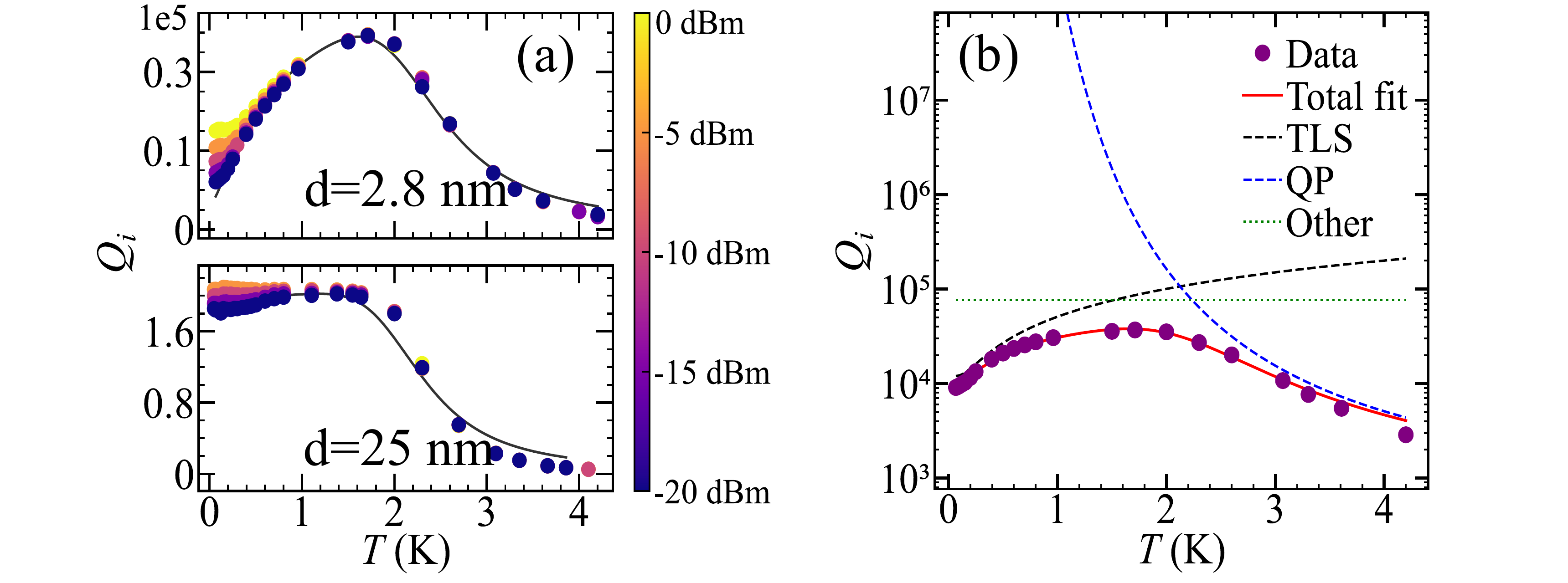}
    \caption{\justifying
    \textbf{Microwave dissipation in NbN resonators:} (a) Internal quality factor $Q_i$ as a function of temperature for 
    representative resonance modes of the $2.8~\mathrm{nm}$ (upper panel) and $25~\mathrm{nm}$ (lower panel) resonators, measured at input powers from $-20$ to $0~\mathrm{dBm}$. The pronounced low-temperature power dependence of the $2.8~\mathrm{nm}$ resonator is consistent with saturation of a TLS-like loss channel, whereas the 
    high-temperature decrease in $Q_i$ results from thermally excited quasiparticles. (b) Inverse internal quality factor $1/Q_i$ of the $2.8~\mathrm{nm}$ resonator at $-20~\mathrm{dBm}$, together with the decomposition into 
    TLS-like, quasiparticle (QP), and residual contributions according to Eq.~\ref{eq:Qloss}. The solid line represents their combined contribution.}
    \label{Figure_5}
\end{figure*}
Beyond the reactive electrodynamic response encoded in the resonant-frequency shift, the internal quality factor $Q_i$ provides complementary information on the dissipative channels that limit the performance of high-kinetic-inductance NbN resonators. Figure~\ref{Figure_5}(a) shows the temperature dependence of $Q_i$ for representative resonators fabricated from the $2.8$ and $25~\mathrm{nm}$ films, measured at several microwave drive powers.\\
The $2.8~\mathrm{nm}$ resonator exhibits a pronounced non-monotonic temperature dependence. At the lowest applied power, $Q_i$ increases from approximately $10^4$ at base temperature to $\sim 3.5\times10^4$ near $1.8~\mathrm{K}$, before decreasing rapidly at higher temperatures. The low-temperature loss is strongly power dependent and is progressively suppressed with increasing microwave drive, consistent with saturation of two-level systems (TLS) associated with dielectric defects in surface oxides and interfaces ~\cite{martinis2005decoherence, muller2019towards,megrant2012planar, mcrae2020materials, pappas2011two, kumar2008temperature, kirsh2017revealing}. Increasing temperature likewise reduces the net resonant TLS absorption through thermal population equalization, accounting for the initial increase in $Q_i$. Because this pronounced 
low-temperature behavior is observed in the $2.8~\mathrm{nm}$ film, the most strongly disordered member of the series, we cannot exclude an additional intrinsic contribution from disorder-induced localized collective modes associated with spatial inhomogeneity of the superconducting state. Such modes 
have been predicted to produce a TLS-like temperature dependence of the microwave quality factor in strongly disordered superconductors~\cite{khvalyuk2026dissipation}. At higher temperatures, however, the increasing population of thermally excited quasiparticles produces a rapid increase of microwave dissipation and the corresponding decrease in $Q_i$.\\

To quantify the competition between the dominant loss channels, we model the measured internal loss as the sum of a TLS-like contribution, a quasiparticle contribution, and an approximately temperature-independent residual term,
\begin{equation}
    \frac{1}{Q_i(T)}
    =
    \frac{1}{Q_{\mathrm{TLS}}(T)}
    +
    \frac{1}{Q_{\mathrm{QP}}(T)}
    +
    \frac{1}{Q_{\mathrm{res}}}.
    \label{eq:Qloss}
\end{equation}
Here, the low-temperature contribution is described using the standard resonant-TLS temperature dependence, whereas the quasiparticle contribution is described within Mattis--Bardeen electrodynamics~\cite{mattis1958theory, zmuidzinas2012superconducting}. We emphasize that the TLS form is used here as the quantitative description of the observed low-temperature loss; given the predicted TLS-like response of localized collective modes in strongly disordered superconductors, the 
temperature dependence alone does not uniquely distinguish between these microscopic origins. The explicit functional forms and details of the fitting procedure are provided in SM.\\

Figure~\ref{Figure_5}(b) shows the resulting decomposition of the measured inverse quality factor $1/Q_i$ for the $2.8~\mathrm{nm}$ resonator at an input power of $-20~\mathrm{dBm}$. At the lowest temperatures, the dissipation is dominated by the TLS-like contribution, consistent with the strong power dependence observed in Fig.~\ref{Figure_5}(a). As the temperature increases, this low-temperature contribution decreases, while quasiparticle loss grows rapidly and eventually becomes the dominant dissipation channel. The competition between these two mechanisms naturally accounts for the nonmonotonic temperature dependence of $Q_i$, while the residual contribution provides an approximately temperature-independent loss background.
The behavior changes markedly for the $25~\mathrm{nm}$ resonator [lower panel of Fig.~\ref{Figure_5}(a)]. Its internal quality factor reaches values of order $10^5$ and exhibits no significant dependence on microwave drive power over the measured range, indicating substantially weaker low-temperature TLS-like dissipation than in the extreme ultrathin limit. Notably, even the next thickness in the series, the 5 nm resonator, shows no discernible low-temperature power dependence characteristic of TLS-like loss. Additional measurements of other resonator modes and film thicknesses are presented in the SM.\\
These measurements complement the frequency-shift analysis by revealing the dissipative consequences of pushing NbN toward the ultrathin, high-kinetic-inductance regime. While reducing the film thickness enhances 
$L_{k,\square}$ and lowers the superfluid stiffness, the strongest increase in low-temperature microwave loss is primarily confined to the extreme ultrathin limit. The fact that this enhanced dissipation appears in the most strongly disordered film is also qualitatively consistent with the possibility of an intrinsic contribution from disorder-induced collective excitations, although the present measurements do not allow this contribution to be separated unambiguously from conventional TLS loss. Thicker NbN films therefore retain 
a favorable combination of large kinetic inductance and comparatively low microwave dissipation, identifying a useful intermediate regime for compact high-impedance superconducting circuits.

\end{document}


\title{\normalfont Supplementary Material\\
\textbf{Power-Law Suppression of Superfluid Stiffness in High-Kinetic-Inductance NbN Films}}

\author{Meenakshi Sharma}
\affiliation{Max Planck Institute for Chemical Physics 
of Solids Nothnitzer Str. 40 01187 Dresden, Germany}
\affiliation{Leibniz Institute for Solid State and Materials Research Dresden (IFW Dresden), Helmholtzstraße 20, 01069 Dresden, Germany}

\author{Hrishikesh Borah}
\affiliation{Max Planck Institute for Chemical Physics 
of Solids Nothnitzer Str. 40 01187 Dresden, Germany}
\affiliation{Technische Universität Dresden, 01062 Dresden, Germany}

\author{Sandeep Singh}
\affiliation{CSIR - National Physical Laboratory, 
Dr. K.S. Krishnan Marg, 110012 - New Delhi, India}

\author{Berit H. Goodge}
\affiliation{Max Planck Institute for Chemical Physics 
of Solids Nothnitzer Str. 40 01187 Dresden, Germany}

\author{Edouard Lesne}
\affiliation{Max Planck Institute for Chemical Physics 
of Solids Nothnitzer Str. 40 01187 Dresden, Germany}

\author{Sandra Nestler}
\affiliation{Leibniz Institute for Solid State and Materials Research Dresden (IFW Dresden), Helmholtzstraße 20, 01069 Dresden, Germany}

\author{Surinder P. Singh}
\affiliation{CSIR - National Physical Laboratory, 
Dr. K.S. Krishnan Marg, 110012 - New Delhi, India}

\author{Haolin Jin}
\affiliation{Max Planck Institute for Chemical Physics 
of Solids Nothnitzer Str. 40 01187 Dresden, Germany}
\affiliation{Technische Universität Dresden, 01062 Dresden, Germany}

\author{Yejin Lee}
\affiliation{Max Planck Institute for Chemical Physics 
of Solids Nothnitzer Str. 40 01187 Dresden, Germany}

\author{Bernd Büchner}
\affiliation{Leibniz Institute for Solid State and Materials Research Dresden (IFW Dresden), Helmholtzstraße 20, 01069 Dresden, Germany}

\author{Uri Vool}
\affiliation{Max Planck Institute for Chemical Physics 
of Solids Nothnitzer Str. 40 01187 Dresden, Germany}
\affiliation{Leibniz Institute for Solid State and Materials Research Dresden (IFW Dresden), Helmholtzstraße 20, 01069 Dresden, Germany}

\maketitle

\setcounter{tocdepth}{1}
\tableofcontents

\section{Thin Film Deposition and Characterization}
Niobium nitride (NbN) thin films with varying thicknesses were deposited on \textit{c}-axis Al$_2$O$_3$ substrates using reactive direct current (DC) magnetron sputtering. The deposition was carried out in an Excel Instruments sputtering system installed at CSIR–National Physical Laboratory, India. Prior to deposition, the substrates were sequentially cleaned in acetone and isopropyl alcohol (IPA) baths under ultrasonic agitation to remove surface contaminants and improve film adhesion.\\
After cleaning, the substrates were loaded into the deposition chamber, which was evacuated to a base pressure of $\sim 10^{-7}$~mbar using a turbomolecular pump. The substrate temperature was then raised to 600~$^\circ$C under PID-regulated control. A 2-inch niobium target with 99.95$\%$ purity was used for sputtering, and the target-to-substrate distance was maintained at 8.5~cm throughout the deposition process. Once the desired base pressure and substrate temperature were achieved, ultra-high-purity N$_2$ and Ar gases, each with 99.999$\%$ purity, were introduced into the chamber at a N$_2$:Ar flow ratio of 1:5. This gas mixture established a steady-state working pressure of $7 \times 10^{-3}$~mbar.
Before initiating NbN film growth, the niobium target was pre-sputtered for 2~min to remove native surface oxides and stabilize the plasma. Following this stabilization step, the films were deposited at a DC sputtering power of 200~W.\\

The surface morphology of the deposited NbN film was further examined by atomic force microscopy (AFM). Figure~\ref{fig:afm} shows the AFM height image of the 2.8~nm-thick NbN film acquired over a 1~$\mu$m $\times$ 1~$\mu$m scan area in tapping mode using a Veeco Instruments system. The image shows a continuous and uniform film with complete substrate coverage. The surface consists of densely packed nanocrystalline grains with no preferred orientation and minimal porosity, in agreement with the nanocrystalline microstructure observed in the cross-sectional STEM image shown in Fig.~1(a) of the main text.\\
The absence of isolated hillocks or abnormal grain growth further indicates homogeneous nucleation and stable growth conditions across the substrate. The total height variation across the scanned region remains below 8~nm, while the root-mean-square surface roughness, defined as $R_q = \sqrt{N^{-1}\sum_{i=1}^{N}(z_i - \bar{z})^2}$, is only 0.2~nm. These results confirm the ultrasmooth morphology of the deposited NbN films.\\
\setcounter{figure}{0}
\renewcommand{\thefigure}{S\arabic{figure}}

\begin{figure}[htbp]
    \centering
    \includegraphics[width=0.3\linewidth]{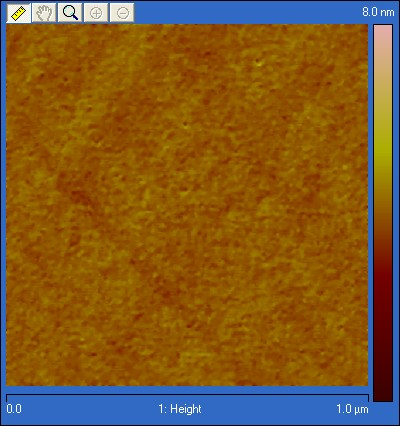}
    \caption{\justifying {AFM topography image of the 2.8~nm NbN film deposited on Al$_2$O$_3$, acquired over a 1~$\mu$m $\times$ 1~$\mu$m scan area in tapping mode. The measured root-mean-square surface roughness is $R_q = 0.2$~nm.}}
    \label{fig:afm}
\end{figure}

\subsection{STEM measurements}
Cross-sectional specimens for investigation by scanning transmission electron microscopy (STEM) were prepared by the standard focussed ion beam (FIB) lift-out method. Rough trenches were cut on a Thermo Fisher Helios G4 UXe plasma FIB. followed by polishing, lifting out, and thinning on a Thermo Fisher Helios 5 CX Ga-FIB. STEM images were collected in annular dark-field geometry on a double aberration-corrected JEOL ARM300F operating at 300 kV with a probe convergence angle of 30 mrad. 

\setcounter{figure}{1}
\renewcommand{\thefigure}{S\arabic{figure}}

\begin{figure}[H]
    \centering
    \includegraphics[width=0.85\linewidth]{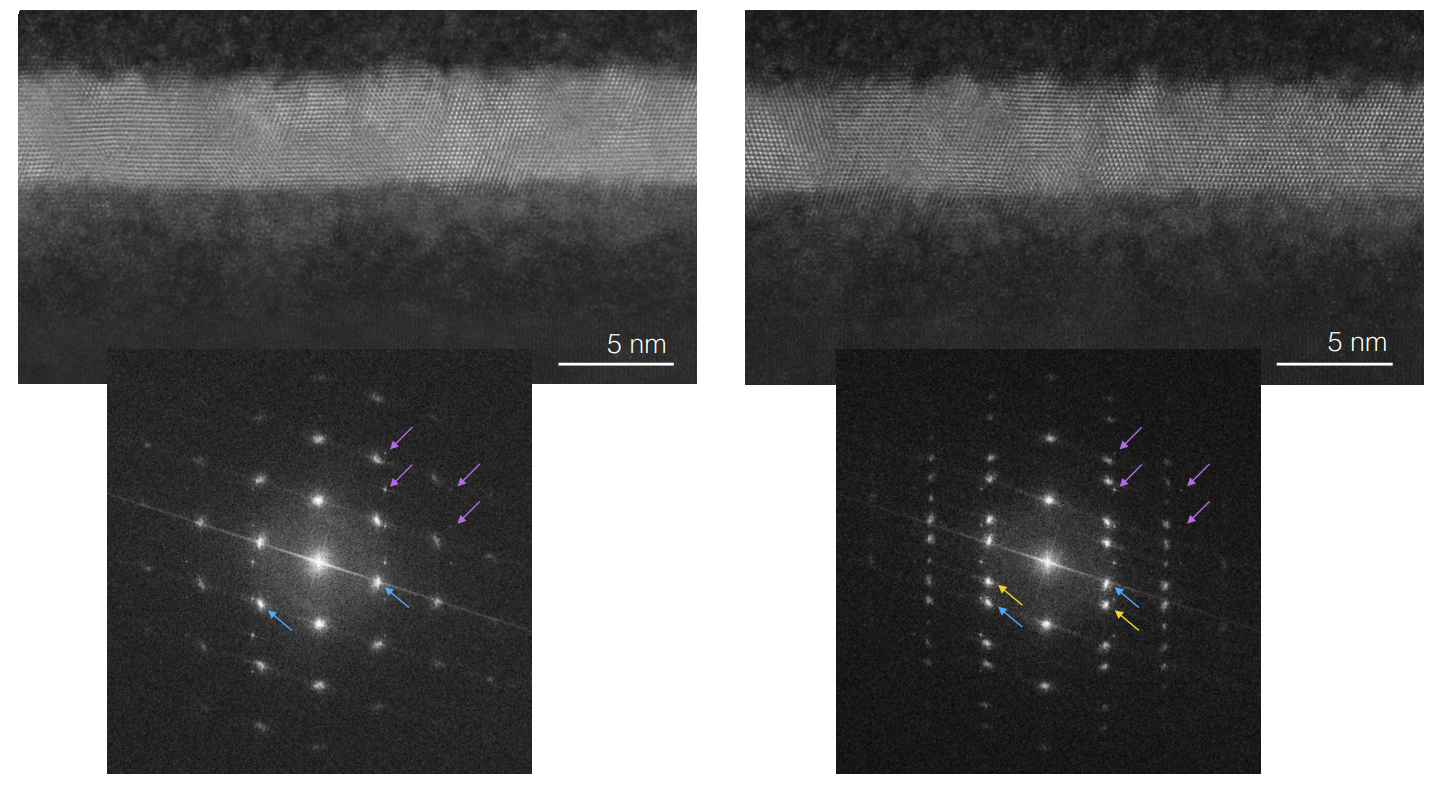}
    \caption{\justifying {Cross-sectional STEM images of the NbN film (top) and their fast Fourier transforms (FFTs, bottom). The FFTs show film peaks (blue) which are much broader than the sharp substrate peaks (purple), indicating reduced long-range structural ordering. In general, consistent FFT patterns indicate that the films are singly oriented relative to the substrate, despite an amorphous interface. Some regions (left) show only one set of film peaks indicating relatively homogeneous film orientation, while other regions (right) show reflected twin orientations (yellow).}}
    \label{fig:stem}
\end{figure}

\section{Device Fabrication}
The NbN resonators were fabricated using an optical lithography followed by plasma reactive ion etching (ICP-RIE) process. The fabrication began by spin coating MIR 701 photoresist onto the deposited NbN films using a two-step spin process. In the first step, the resist was spun at 4500~rpm with an acceleration of 1000~rpm/s for 45~s to define the main resist thickness. This was followed by a second spin at 10000~rpm for 10~s to remove excess residual resist and improve coating uniformity. The samples were then soft-baked at 90~$^\circ$C for 90 seconds, yielding a nominal photoresist thickness of approximately 1.8~$\mu$m. The Patterning of the simulated design was carried out using a direct laser writing system, MLA 100 (Heidelberg Instruments), with an exposure dose of 350~mJ/cm$^2$. After exposure, the sample was post-baked at 115~$^\circ$C for 60 sec and subsequently developed for 2~min. Prior to etching, a brief oxygen plasma de scum was performed to remove residual organic or lithographic residues from the exposed regions.
The final resonator geometry was then etched into the NbN film by ICP-RIE using a PlasmaPro 100 Cobra system from Oxford Instruments. The etching was performed with a table RF power of 30~W and an ICP RF power of 500~W. The etch recipe was optimized with respect to both the NbN film thickness and the Al$_2$O$_3$ substrate. The best etching selectivity between the optical resist and the NbN thin film was obtained using a CF$_4$/Ar gas mixture with flow rates of 90~sccm and 10~sccm, respectively, at an operating pressure of 50~mTorr.
For device fabrication, a controlled overetch of approximately 4-5~nm was typically used. This slight overetch provides sufficient process margin to compensate for small variations in film thickness, etch rate, and local plasma uniformity, ensuring complete removal of NbN from the exposed areas. Complete clearing of the unwanted NbN is important for avoiding residual conductive paths or etched residues that could introduce microwave loss and degrade the resonator performance. At the same time, the overetch was kept limited to minimize unnecessary exposure of the Al$_2$O$_3$ substrate to the plasma, since excessive substrate etching or plasma damage can adversely affect the device quality factor.
After etching, the remaining photoresist was removed by immersing the sample in TechniStrip P1330 and ultrasonically agitating it for 5~min. The sample was then further cleaned in acetone and isopropyl alcohol (IPA) baths under ultrasonic agitation for 5~min each. Finally, the etched depth was verified using a profilometer to confirm complete NbN removal and to assess the reproducibility of the etching process.\\

\section{Transport measurements}
The electrical characterization of the deposited nanofilms was carried out by measuring Hall bar devices corresponding to each film thickness. For every film in the thickness series, the sheet resistance \(R_{\square}(T)\) and the superconducting transition temperature \(T_c\) were extracted using a standard four-probe geometry, in which a DC current was sourced along the Hall bar length and the resulting longitudinal voltage drop was detected across two side contacts, thereby eliminating contributions from contact and lead resistances.
The samples were thermally cycled in a Quantum Design Physical Property Measurement System (PPMS). The \(R\)-\(T\) characteristics were acquired using a source-measure unit, Keysight B2912A, operating in DC mode, with the sourced current kept at \(\leq \SI{1}{\micro\ampere}\) to minimize Joule heating and self-field effects in the superconducting state. The measurements spanned the temperature range of \(\SIrange{2}{300}{\kelvin}\), with the sample temperature regulated and stabilized by the PPMS and monitored using its calibrated thermometry.\\
The sheet resistance was obtained from the measured resistance through
\[
R_{\square} = \frac{\rho}{d} = R\left(\frac{w}{l}\right),
\]
where \(w\) and \(l\) are the width and length of the Hall bar channel, respectively. Each film was measured during both a cooling sweep and a subsequent warming sweep to verify reproducibility and rule out heating artifacts; the two sweeps agreed closely for all films.
Figure~\ref{fig:RT} shows \(R_{\square}(T)\) for the full thickness series, measured from room temperature down through the superconducting transition. In the normal state, all films exhibit a characteristic resistance maximum above \(T_c\). Above this maximum, \(R_{\square}\) decreases monotonically with increasing temperature up to \(\SI{300}{\kelvin}\), a hallmark of strongly disordered conductors in which normal-state transport is governed by disorder rather than phonon scattering, consistent with the dirty-limit regime of these films~\cite{chockalingam2008superconducting, Kern2024}. Below the maximum, \(R_{\square}(T)\) decreases monotonically before dropping sharply to zero at the transition.
The normal-state sheet resistance \(R_N\), defined as the value of \(R_{\square}\) at the resistance maximum, increases monotonically as the film thickness is reduced, rising from \(\SI{128.7}{\ohm}/\square\) for the \(\SI{25}{\nano\meter}\) film to \(\SI{1686.4}{\ohm}/\square\) for the \(\SI{2.8}{\nano\meter}\) film (Table~I, main text). This thickness dependence indicates progressively stronger disorder in thinner films, where enhanced surface and grain-boundary scattering reduce the electron mean free path \(\ell\) as the film thickness approaches the nanocrystalline grain size. Consequently, the thinnest films lie deepest in the dirty-limit regime~\cite{chockalingam2008superconducting}.
The superconducting transition temperature was determined as
\[T_c = \frac{T_{10\%} + T_{90\%}}{2},\]
where \(T_{10\%}\) and \(T_{90\%}\) are the temperatures at which \(R_{\square}\) falls to \(10\%\) and \(90\%\) of \(R_N\), respectively~\cite{sharma2026berezinskii}. The extracted \(T_c\) decreases systematically from \(\SI{11.7}{\kelvin}\) at \(\SI{25}{\nano\meter}\) to \(\SI{7.72}{\kelvin}\) at \(\SI{2.8}{\nano\meter}\), consistent with the thickness driven suppression of \(T_c\) \cite{chockalingam2008superconducting, joshi2018superconducting}. The resulting \(R_N\) and \(T_c\) values were used to evaluate the zero-temperature sheet kinetic inductance \(L_{k,\square}(0)\) for each film via Eq.~(3) of the main text, as detailed in the Simulation section.\\
\setcounter{figure}{2}
\renewcommand{\thefigure}{S\arabic{figure}}
\begin{figure}[htbp]
\centering
\includegraphics[width=0.65\linewidth]{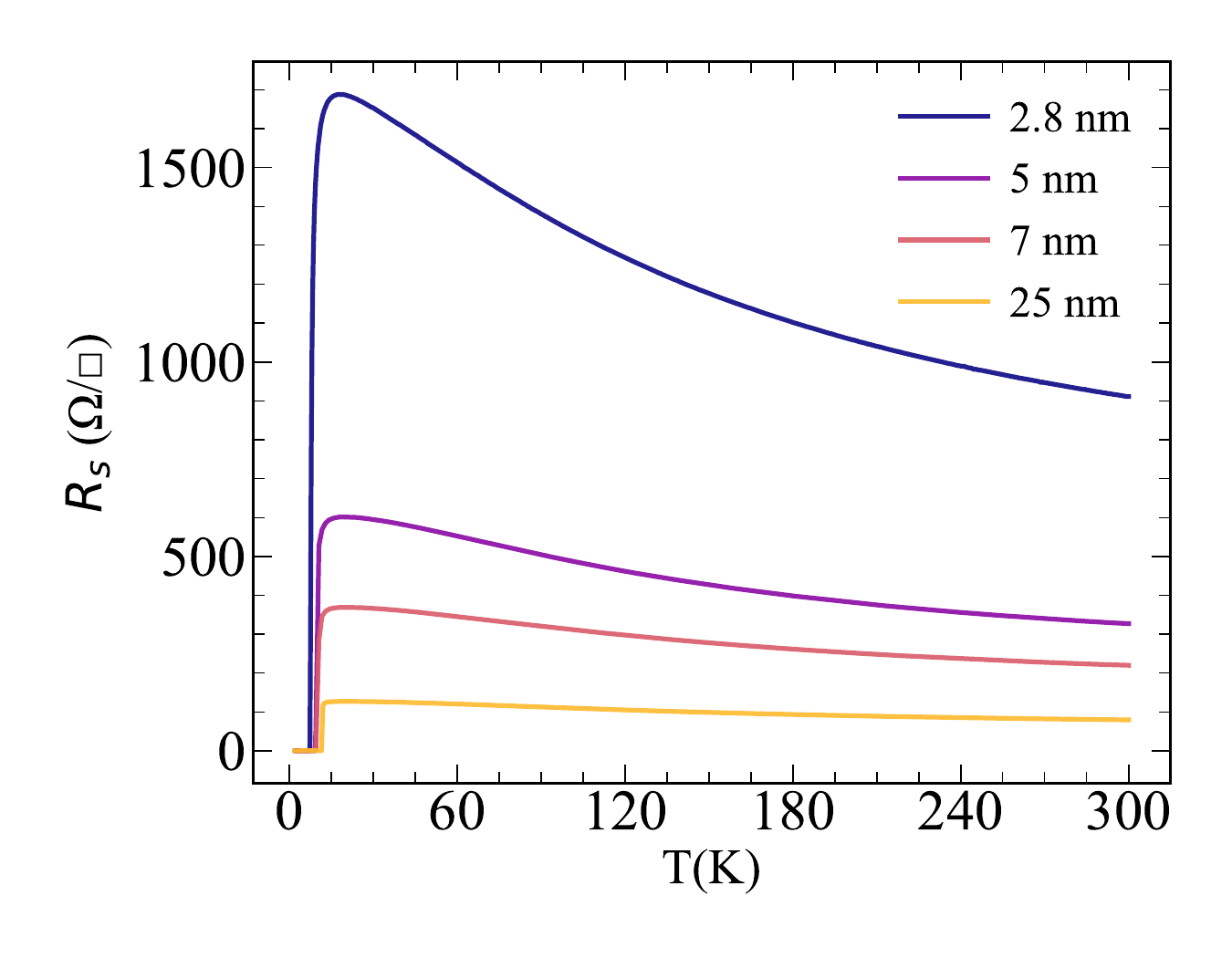}
\caption{\justifying {Sheet resistance $R_\square$ as a function of temperature for NbN films with thicknesses ranging from 2.8~nm to 25~nm. All films exhibit a resistance maximum above $T_c$ followed by a sharp superconducting transition. The resistance maximum is most pronounced in the thinnest films, reflecting enhanced weak localization corrections.}}
\label{fig:RT}
\end{figure}
Moreover, while Tc is defined from the resistive midpoint as above, the 2.8 nm film exhibits a markedly broadened transition with an extended fluctuation regime, which we analyze below to identify the BKT transition.

\subsection{Observation of Berezinskii-Kosterlitz-Thouless transition}
For the thinnest (2.8~nm) film, the superconducting transition is broadened
and develops a resistive tail across an extended fluctuation regime above
$T_c$ [Fig.~1(b) of the main text]. To characterize this regime
quantitatively, we analyze the excess paraconductivity
\begin{equation}
\Delta\sigma(T) = \frac{1}{R_\Box(T)} - \frac{1}{R_{\Box N}(T)},
\end{equation}
where the normal-state sheet resistance is described by
$R_{\Box N}(T) = A + B/T + C/T^2 + D/T^3$, fitted over the 14--19~K interval
well above the transition.

Far from the transition, the data are well described by the two-dimensional
Aslamazov--Larkin (AL) paraconductivity~\cite{aslamasov1968influence}, which
captures Gaussian amplitude fluctuations of the order parameter,
\begin{equation}
\Delta\sigma_{\rm AL}(T) = \frac{e^2/(16\hbar d)}{s\,\sinh(\epsilon/s)},
\qquad \epsilon = \ln(T/T_{\rm AL}),
\end{equation}
yielding $T_{\rm AL} = 7.721 \pm 0.005$~K. Closer to the transition the AL
form deviates from the data, and the Halperin--Nelson (HN)
form~\cite{kosterlitz1973ordering,nelson1977universal} takes over, reflecting
the exponential divergence of the superconducting correlation length as free
vortices bind into pairs~\cite{sharma2026berezinskii},
\begin{equation}
\Delta\sigma_{\rm HN}(T) = \frac{1}{R_{\Box N}(T)}\,
a_{\rm HN}\,\sinh\!\left[ b_{\rm HN}
\sqrt{\frac{T_{\rm HN}}{T - T_{\rm HN}}} \right],
\end{equation}
yielding $T_{\rm HN} \equiv T_{\rm BKT} = 6.959 \pm 0.026$~K. The
AL\,$\rightarrow$\,HN crossover near the transition [inset of Fig.~1(b),
main text], together with the gap
$T_{\rm AL} - T_{\rm BKT} \simeq 0.76$~K, identifies the BKT regime and
reflects the disorder-suppressed superfluid stiffness expected in this
two-dimensional limit.

\section{Electromagnetic simulation}

\subsection{Kinetic Inductance and Simulation Setup}
In a superconducting transmission line or resonator, the total inductance per unit length has two contributions: a geometric part $L_g$, arising from the magnetic energy stored in the electromagnetic field surrounding the conductor, and a kinetic part $L_k$, arising from the kinetic energy of the Cooper pair condensate. In the dirty limit (mean free path $\ell \ll \xi_0$, where $\xi_0$ is the BCS coherence length) relevant to our NbN films, the sheet kinetic inductance ($L_{k,\text{per length}} = \frac{L_{k,\square}}{w}$) is given by the Mattis-Bardeen relation~\cite{Zmuidzinas2012}:
\begin{equation}
    L_{k,\square}(T) = \frac{\hbar R_\square}{\pi\Delta(0)} 
    \frac{1}{\tanh\left[\Delta(0)/(2k_BT)\right]},
    \label{eq:Lk_MB}
\end{equation}
where \(R_\square\) denotes the normal-state sheet resistance measured just above the superconducting transition temperature \(T_C\), and \(\Delta(0) \approx 1.764\,k_B T_C\) is the superconducting energy gap at zero temperature. As the film thickness decreases, $R_\square$ grows rapidly (Table~I of the main text), causing 
$L_{k,\square}$ to increase correspondingly. As a result, in our ultrathin nanocrystalline NbN films, $L_{k,\square}$ becomes the dominant contribution to the total inductance, and the resonator's electromagnetic energy is stored primarily in the form of kinetic energy of the condensate. At base temperature, $\tanh[\Delta(0)/(2k_BT)] \to 1$, 
so Eq.~\eqref{eq:Lk_MB} reduces to $L_{k,\square}(0) = \hbar R_\square / \pi\Delta(0)$, which was used to extract the sheet kinetic inductance for each film from the transport measurements. Across the thickness series, $L_{k,\square}(0)$ rises from $14~$pH/$\square$ at 25~nm to $300~$pH/$\square$ at 2.8~nm, as summarized in Table~I of the main text.\\
Electromagnetic simulations of the NbN coplanar-waveguide (CPW) resonators were performed in Sonnet, which solves Maxwell's equations for the classical electromagnetic response of a planar structure. Since kinetic inductance is a quantum-mechanical property of the superconducting condensate, it is not captured directly by the field solver. Instead, it was incorporated by assigning a finite sheet kinetic inductance, \(L_{k,\square}\), to the NbN layer through the reactive part of the surface impedance in the Sonnet material definition.
The resonators have a center-conductor width \(w = 10~\mu\mathrm{m}\) and gap \(s = 6~\mu\mathrm{m}\), and are fabricated on a \(500~\mu\mathrm{m}\)-thick \(c\)-axis sapphire substrate with relative permittivity \(\varepsilon_r = 9.4\). The simulation setup is shown in the upper panel of Fig.~\ref{fig:sonnet_combined}.\\
To isolate the effect of kinetic inductance on the resonance frequency, two sets of simulations were performed: one including only the geometric inductance, \(L_g\), corresponding to \(L_k = 0\), and another including the total inductance, \(L_g + L_k\). For a distributed transmission-line resonator of length \(\ell\), the fundamental resonance frequency is

\begin{equation}
    f_0 = \frac{1}{2\ell\sqrt{(L_g + L_k)C}},
    \label{eq:resonance_freq}
\end{equation}

where \(L_g\), \(L_k\), and \(C\) are the geometric inductance, kinetic inductance, and capacitance per unit length, respectively. In the limit where kinetic inductance dominates, \(L_k \gg L_g\), the resonance frequency scales as

\begin{equation}
    f_0 \propto \frac{1}{\sqrt{L_k}}.
\end{equation}
Thus, the resonance frequency provides a direct probe of the kinetic inductance and, consequently, of the superfluid density. This forms the physical basis for using the fractional frequency shift, \(\delta f/f\), as a sensitive electrodynamic probe throughout the main text.\\
For the \(2.8~\mathrm{nm}\) NbN film, including kinetic inductance lowers the simulated resonance frequency by approximately an order of magnitude relative to the geometric-only case, as shown in Fig.~\ref{fig:sonnet_combined}(a,b). To verify the expected scaling, the simulation was repeated over a range of \(L_{k,\square}\) values for the same CPW geometry. The resulting fundamental frequency, plotted as a function of \(L_{k,\square}\) in Fig.~\ref{fig:sonnet_combined}(c), confirms the expected scaling,
\begin{equation}
    f_0 \propto \frac{1}{\sqrt{L_{k,\square}}},
\end{equation}

consistent with the kinetic-inductance-dominated regime~\cite{Zmuidzinas2012}.\\


\subsection{Resonator Design Across the Thickness Series}
In our design, each chip hosts two resonators: a quarter-wave (\(\lambda/4\)) resonator with one end shorted to ground, and a half-wave (\(\lambda/2\)) resonator with open terminations at both ends - both capacitively coupled to a shared \(50~\Omega\) feedline in a hanger-type geometry. Both resonators incorporate a meandering layout to achieve the required electrical length while maintaining compact chip dimensions, and their spatial separation minimizes unwanted electromagnetic coupling between them. Accordingly, the resonance condition Eq.~\eqref{eq:resonance_freq} carries a prefactor \(1/(2\ell)\) for the half-wave resonator and \(1/(4\ell)\) for the quarter-wave resonator.
To quantify how strongly each resonator is governed by kinetic rather than geometric inductance, we define the kinetic inductance fraction
\begin{equation}
\alpha = \frac{L_{k,\square}}{L_{k,\square} + L_{g,\square}},
\label{eq:alpha}
\end{equation}
where \(\alpha \rightarrow 1\) indicates that the resonator operates deep in the kinetic-inductance-dominated regime. For our CPW geometry with \(w = 10~\mu\mathrm{m}\) and \(s = 6~\mu\mathrm{m}\), standard conformal mapping~\cite{simons2001coplanar} gives a geometric sheet inductance \(L_{g,\square} = 4.39~\mathrm{pH}/\square\), fixed by the cross-sectional geometry and independent of film thickness. The corresponding capacitance per unit length, evaluated using the equivalent anisotropic permittivity of \(c\)-axis sapphire \(\varepsilon_{\rm eff} = \sqrt{\varepsilon_\perp \varepsilon_\parallel} \approx 10.4\) with \(\varepsilon_\perp = 9.4\) and \(\varepsilon_\parallel = 11.6\), is \(C \approx 150~\mathrm{pF/m}\), this value enters the resonance frequency through Eq.~\eqref{eq:resonance_freq}. Table~\ref{tab:alpha} summarizes \(\alpha\) for each film in the series.

\begin{table}[htbp]
\centering
\caption{Sheet kinetic inductance \(L_{k,\square}\) and kinetic inductance fraction \(\alpha\) for each NbN film thickness.}
\label{tab:alpha}
\begin{tabular}{ccc}
\hline\hline
$d$ (nm) & $L_{k,\square}$ (pH/$\square$) & $\alpha$ \\
\hline
2.8  & 300 & 0.986 \\
5    & 85  & 0.951 \\
7    & 49  & 0.918 \\
25   & 14  & 0.761 \\
\hline\hline
\end{tabular}
\end{table}

\noindent Even the thickest \(25~\mathrm{nm}\) film remains strongly kinetic-inductance dominated \((\alpha = 0.761)\), confirming that all resonators in the series operate in the kinetic-inductance-dominated regime. In the limit \(\alpha \approx 1\), the fractional frequency shift directly reflects the relative change in kinetic inductance via Eq.~(2) of the main text:
\begin{equation}
\frac{\delta f(T)}{f} \approx -\frac{\alpha}{2}\frac{\delta\Theta(T)}{\Theta(0)},
\label{eq:freq_shift}
\end{equation}
making \(\delta f/f\) a quantitatively reliable probe of the condensate electrodynamics across the full thickness series.\\
The physical resonator length was adjusted individually for each film to bring the fundamental resonance frequency into the \(2\)-\(13~\mathrm{GHz}\) measurement bandwidth of the VNA. For example, at the unadjusted design length, the \(2.8~\mathrm{nm}\) film would resonate near \(\sim 1.3~\mathrm{GHz}\) for the \(\lambda/2\) resonator and \(\sim 0.66~\mathrm{GHz}\) for the \(\lambda/4\) resonator, both below the \(2~\mathrm{GHz}\) lower limit of our VNA. Each resonator was therefore designed individually using the Sonnet-extracted \(f_0\) versus \(L_{k,\square}\) relationship [Fig.~\ref{fig:sonnet_combined}(c)] to place the fundamental mode within the accessible bandwidth.\\
The coupling of each resonator to the feedline was likewise designed in Sonnet. The loaded quality factor of a hanger-type resonator separates into an internal contribution \(Q_i\), set by dissipation within the resonator, and a coupling contribution \(Q_c\), set by energy loss to the feedline, related by \(Q_l^{-1} = Q_i^{-1} + Q_c^{-1}\). In the lumped-element representation of Fig.~2(a) of the main text, the distributed kinetic and geometric inductance of the CPW resonator combine into a single inductance \(L_k + L_g\), the distributed capacitance appears as \(C\), and the coupling to the feedline is represented by the coupling capacitor \(C_c\), which sets \(Q_c\). For each film, the coupling-gap dimensions were adjusted to target \(Q_c \sim 10^4\), comparable to the internal quality factor so that the resonators sit near critical coupling. The \(50~\Omega\) feedline was included explicitly in the simulation geometry to capture the electromagnetic environment seen by the resonator, and the simulated \(S_{21}\) response was fitted with the same circle-fitting procedure applied to the experimental data, ensuring that \(Q_c\) is extracted consistently between simulation and experiment.

\setcounter{figure}{3}
\renewcommand{\thefigure}{S\arabic{figure}}

\begin{figure}[htbp]
    \centering

    \includegraphics[
        width=\textwidth,
        trim=90 30 90 50,
        clip
    ]{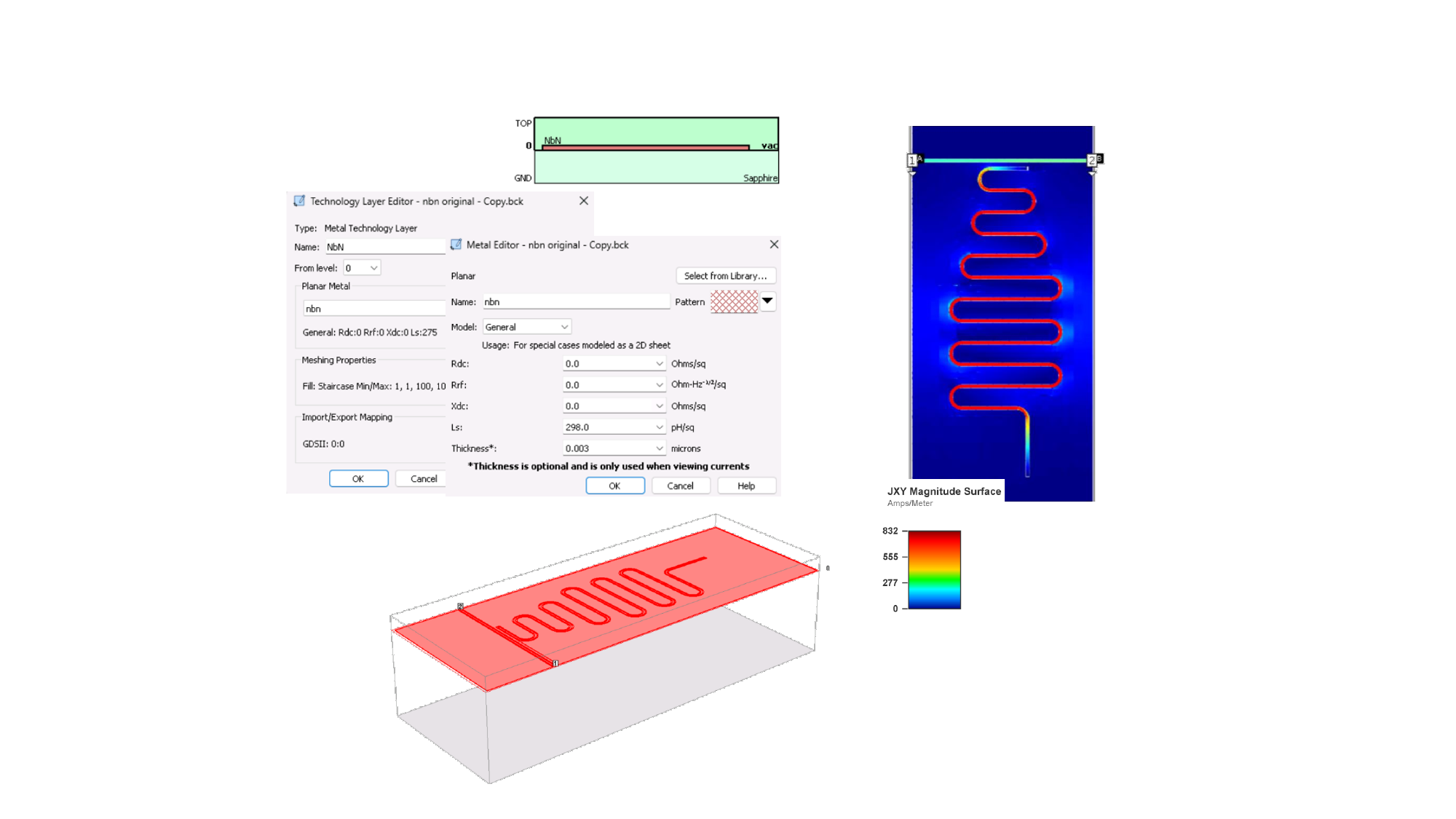}

    \vspace{0.6cm}

    \begin{subfigure}[t]{0.32\textwidth}
        \centering
        \includegraphics[
            width=\linewidth,
            height=5.8cm,
            keepaspectratio
        ]{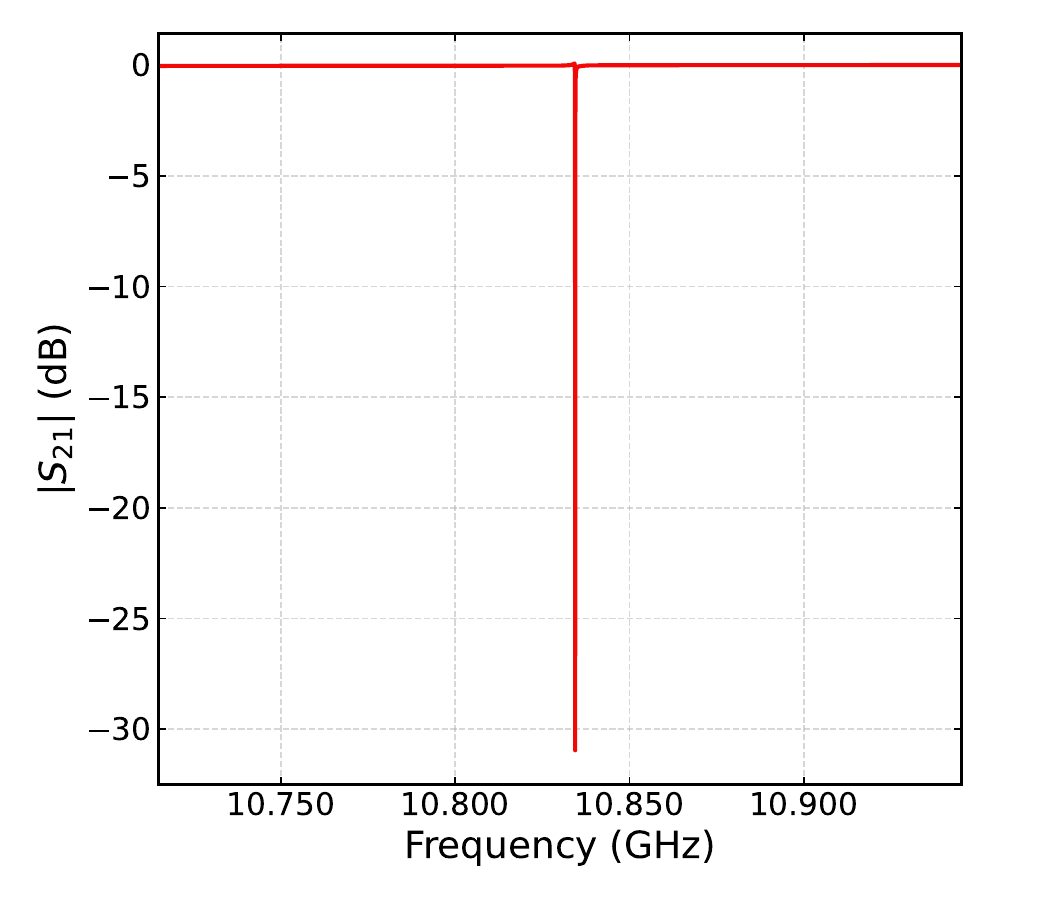}
        \caption{No kinetic inductance 
        ($L_{k,\square}=0$~pH/$\square$).}
    \end{subfigure}
    \hfill
    \begin{subfigure}[t]{0.32\textwidth}
        \centering
        \includegraphics[
            width=\linewidth,
            height=5.8cm,
            keepaspectratio
        ]{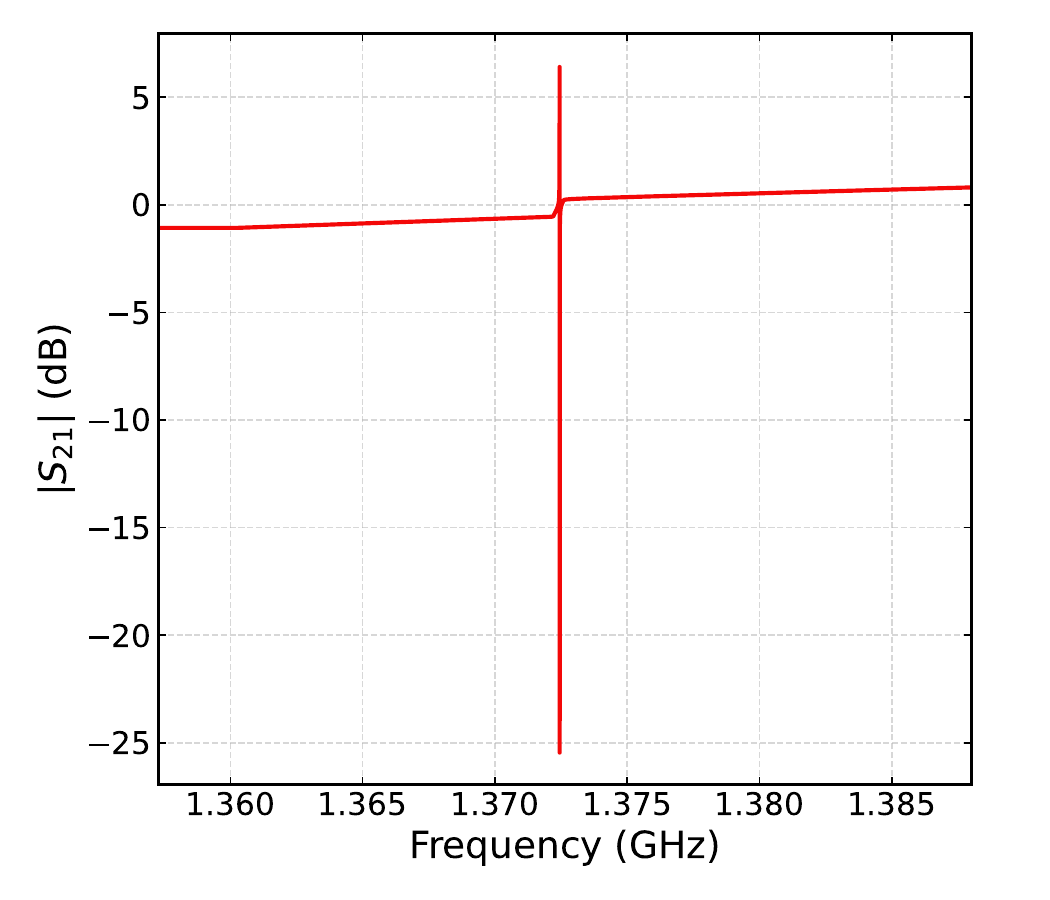}
        \caption{With $L_{k,\square} = 298$~pH/$\square$.}
    \end{subfigure}
    \hfill
    \begin{subfigure}[t]{0.32\textwidth}
        \centering
        \includegraphics[
            width=\linewidth,
            height=5.8cm,
            keepaspectratio
        ]{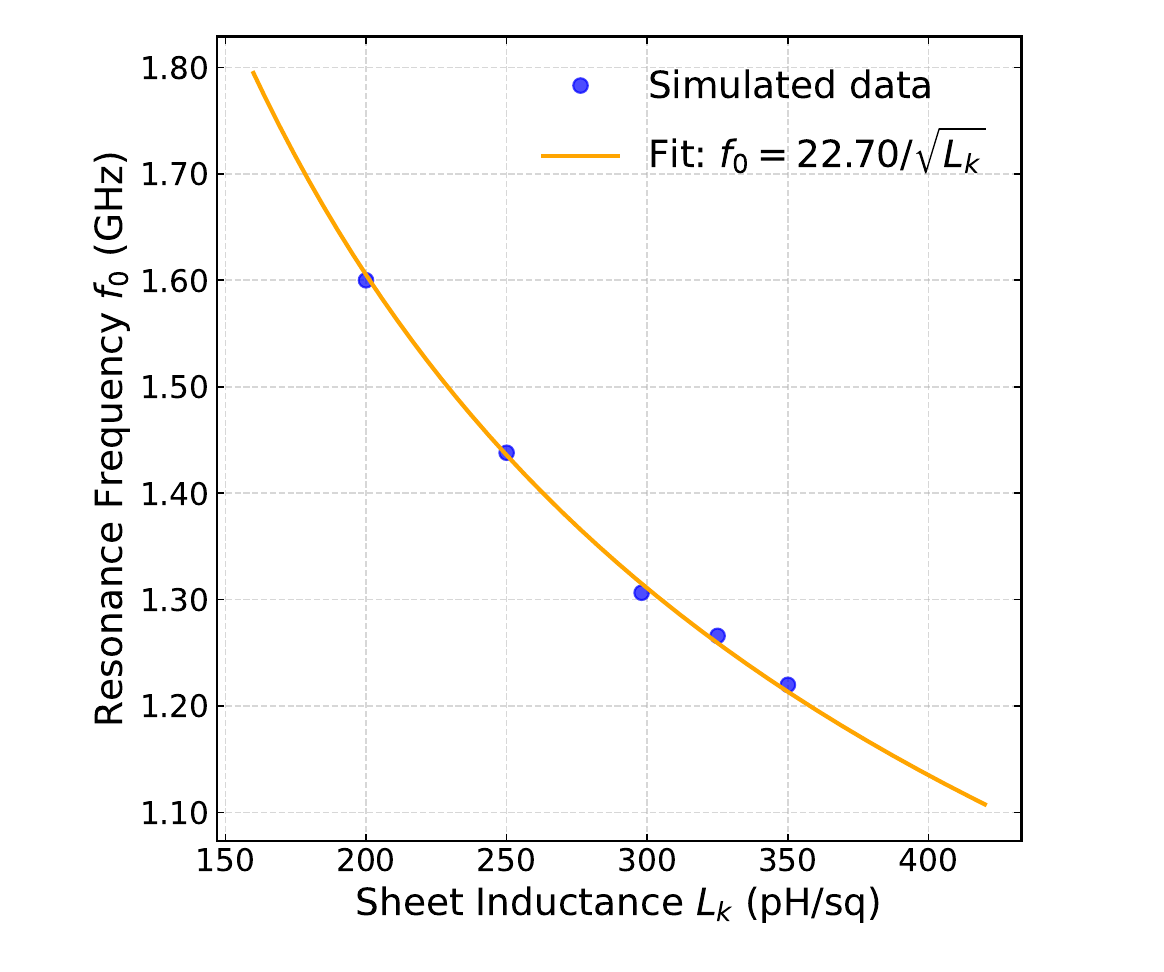}
        \caption{$f_0$ vs.\ $L_{k,\square}$.}
    \end{subfigure}

\caption{\justifying {Electromagnetic simulation of the NbN CPW resonators (\(w = 10~\mu\mathrm{m}\), \(s = 6~\mu\mathrm{m}\)) on \(c\)-axis sapphire with in-plane permittivity \(\varepsilon_\perp = 9.4\), out-of-plane permittivity \(\varepsilon_\parallel = 11.6\), and thickness \(500~\mu\mathrm{m}\). On the right side, the Simulated distribution of the surface current magnitude \(|\mathbf{J}|\) (\(J_{XY}\) surface, \(\mathrm{A/m}\)) for the \(\lambda/2\) resonator at resonance is shown. The half-wave boundary conditions produce current nodes, shown in blue, at both open ends and a current antinode, shown in red, at the midpoint of the line, with a peak value of \(\sim 832~\mathrm{A/m}\). Since the resonators are kinetic-inductance dominated, the high-current region corresponds to where the electromagnetic energy is stored as kinetic energy of the condensate; the enhanced current along the conductor edges reflects the characteristic CPW edge-current distribution. (a)~Simulated \(|S_{21}|\) with no kinetic inductance, showing the purely geometric resonance frequency. (b)~Simulated \(|S_{21}|\) incorporating \(L_{k,\square} = 298~\mathrm{pH}/\square\), demonstrating an order-of-magnitude suppression of \(f_0\). (c)~Fundamental resonance frequency $f_0$ as a function of sheet kinetic 
inductance $L_{k,\square}$ for a fixed $\lambda/2$ CPW geometry, demonstrating the expected $f_0 \propto 1/\sqrt{L_{k,\square}}$ scaling.}}
    \label{fig:sonnet_combined}
\end{figure}

\section{Microwave measurements}
All devices were measured in a BlueFors LD400 dilution refrigerator at a base temperature of approximately \(40~\mathrm{mK}\) at the sample stage. Microwave transmission was recorded with a Keysight P9373A Streamline Series USB vector network analyzer using an IF bandwidth of \(200~\mathrm{Hz}\) and a minimum of 20 averages per trace. On the input line, the microwave signal was attenuated by a total of \(39~\mathrm{dB}\) distributed across the temperature stages of the refrigerator - \(-3~\mathrm{dB}\) at \(300~\mathrm{K}\), \(-10~\mathrm{dB}\) at \(50~\mathrm{K}\), \(-10~\mathrm{dB}\) at \(4~\mathrm{K}\), \(-6~\mathrm{dB}\) at \(80~\mathrm{mK}\), and \(-10~\mathrm{dB}\) at \(40~\mathrm{mK}\) - which both thermalizes the line and suppresses room-temperature Johnson noise reaching the device. On the output side, the transmitted signal was first amplified by a HEMT amplifier at the \(4~\mathrm{K}\) stage and then by a room-temperature low-noise amplifier before detection at the VNA. All measurements were carried out at nominally zero applied magnetic field. Powers quoted throughout this work refer to the VNA output; owing to the \(39~\mathrm{dB}\) of input-line attenuation, the power at the device is lower by approximately the same amount.
\setcounter{figure}{4}
\renewcommand{\thefigure}{S\arabic{figure}}

\begin{figure}[htbp]
    \centering
    \includegraphics[width=\textwidth]{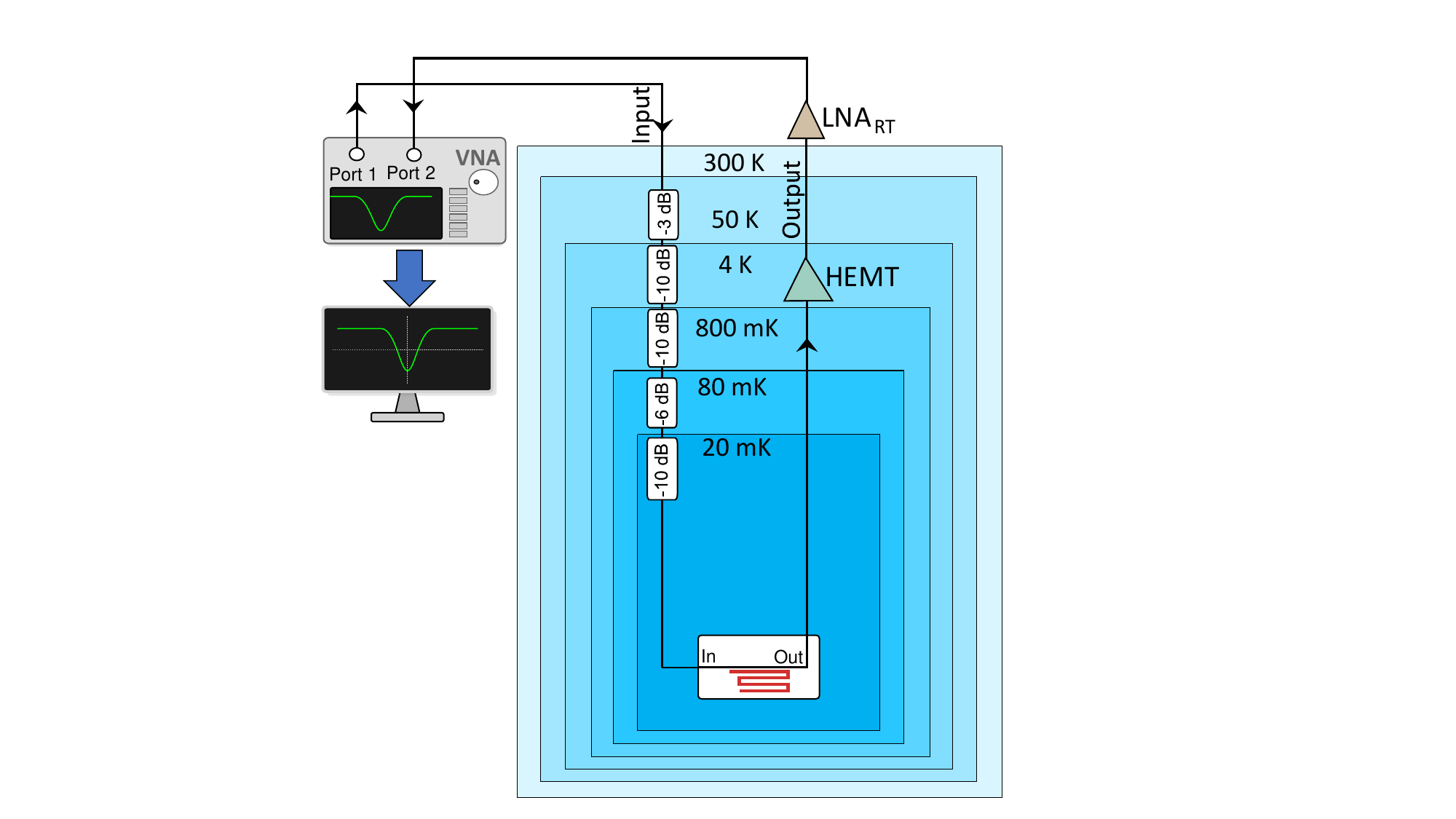}
    \caption{\justifying{Schematic of the dilution refrigerator wiring and microwave measurement setup, including the input attenuation and output amplification stages.}}
    \label{fig:schematic}
\end{figure}

\subsection{Methodology}
Microwave transmission measurements were performed by measuring the complex scattering parameter $S_{21}$ as a function of frequency using a vector network analyzer (VNA). Each resonance was measured over a narrow frequency window centered on the resonance frequency $f_0$, with sufficient averaging. The input power at the device was varied between $-20$~dBm and $5$~dBm at the VNA output port during the measurements. To extract the resonator parameters from the measured $S_{21}$ data, we employed a circle-fitting procedure in the complex plane~\cite{probst2015efficient}. The raw transmission data were first corrected for the background arising from cable delays, impedance mismatches, and amplifier response. This was accomplished by fitting a linear baseline to the off-resonance portions of the complex $S_{21}$ data and normalizing accordingly. The background-corrected data trace a circle in the complex plane, as shown in the inset of Fig.~\ref{fig:s21_combined}, whose diameter and position encode the resonator's coupling and internal losses.
The resonance frequency $f_0$, the loaded quality factor $Q_l$, and the coupling quality factor $Q_c$ were extracted by fitting the background-corrected data to the standard input-output model of a resonator coupled to a transmission line~\cite{probst2015efficient, khalil2012analysis}:
\begin{equation}
    S_{21}(f) = 1 - \frac{Q_l / Q_c}{1 + 2iQ_l\left(\frac{f - f_0}{f_0}\right)}.
    \label{eq:S21_model}
\end{equation}
\setcounter{figure}{5}
\renewcommand{\thefigure}{S\arabic{figure}}

\begin{figure}[htbp]
    \centering
    \includegraphics[width=\textwidth]{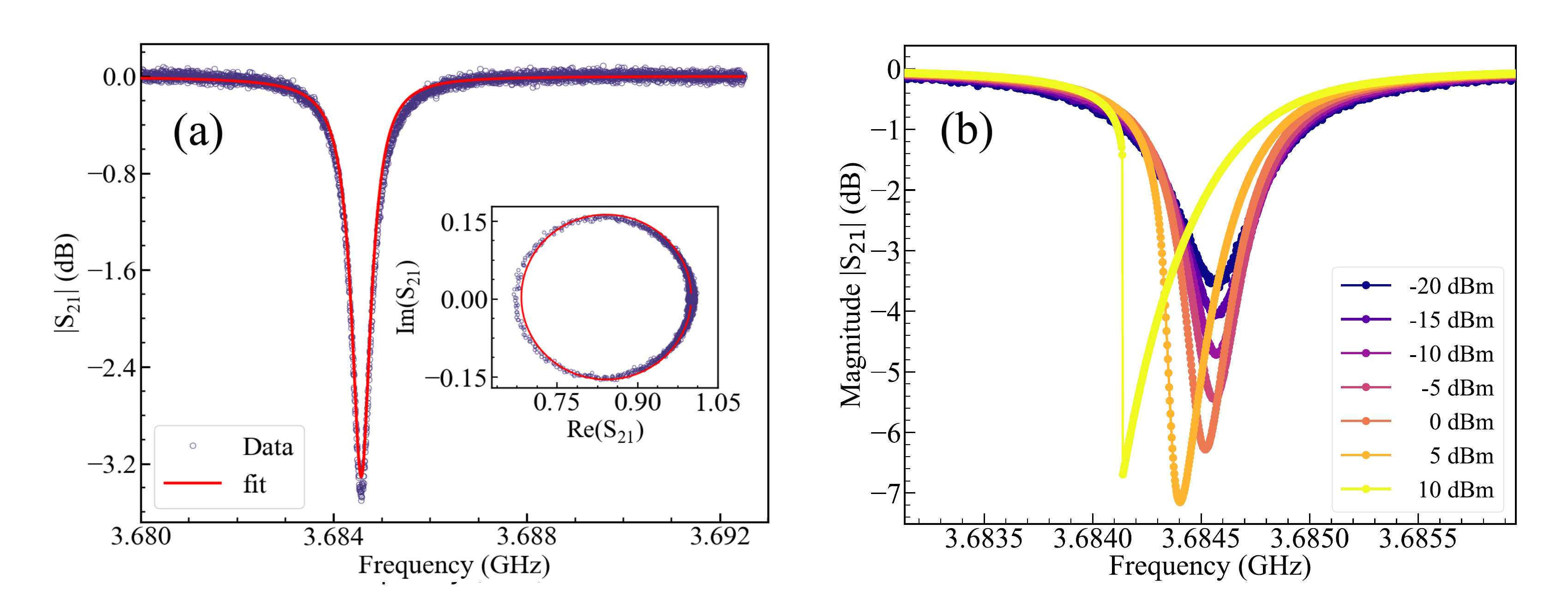} 
    \caption{\justifying {Microwave characterization of a 2.8~nm NbN resonator. 
    (a) Representative $S_{21}$ magnitude spectrum near $f_0 \approx 3.684$~GHz 
    (blue circles) with Lorentzian fit (red line). Inset: circle fit in the complex 
    $S_{21}$ plane. The fit yields $Q_i \approx 1.38 \times 10^4$, 
    $Q_l \approx 9.6 \times 10^3$, and $Q_c \approx 2.0 \times 10^4$. 
    (b) Power-dependent $S_{21}$ spectra near 3.684~GHz for drive powers 
    from $-20$~dBm to $+10$~dBm, showing the onset of Duffing nonlinearity.}}
    \label{fig:s21_combined}
\end{figure}
The internal quality factor was then obtained from the relation $1/Q_i = 1/Q_l - \mathrm{Re}(1/Q_c)$, where the complex $Q_c$ accounts for any impedance asymmetry in the coupling. All fits were performed using a nonlinear least-squares algorithm applied to the real and imaginary parts of the complex $S_{21}$ simultaneously. The fitting was repeated independently for each temperature point and each input power level, yielding the temperature and power-dependent $Q_i$ and $f_0$ reported in this work.\\

At elevated microwave powers, the resonators exhibit a pronounced nonlinear response characteristic of the Duffing oscillator regime. As shown in Fig.~\ref{fig:s21_combined}, the resonance lineshape becomes increasingly asymmetric with increasing drive power, eventually developing the characteristic bifurcation feature of a driven nonlinear oscillator. This behavior arises from the power-dependence of the kinetic inductance, which introduces a Kerr-type photon--photon interaction into the resonator Hamiltonian~\cite{Zmuidzinas2012}. The extracted single-photon Kerr coefficient is $K/2\pi \approx -0.7$~mHz/photon, where the negative sign indicates that increasing photon occupation shifts the resonance to lower frequencies, consistent with the softening of the kinetic inductance under increasing current. All quantitative analysis of the quality factors and frequency shifts reported in this work was therefore performed using data acquired at low input powers, where the resonator operates in the linear regime and Eq.~\eqref{eq:S21_model} remains valid.

\section{Additional data from other modes}
\begin{figure}[htbp]
    \centering
    \begin{subfigure}[t]{0.48\textwidth}
        \centering
        \includegraphics[width=\linewidth]{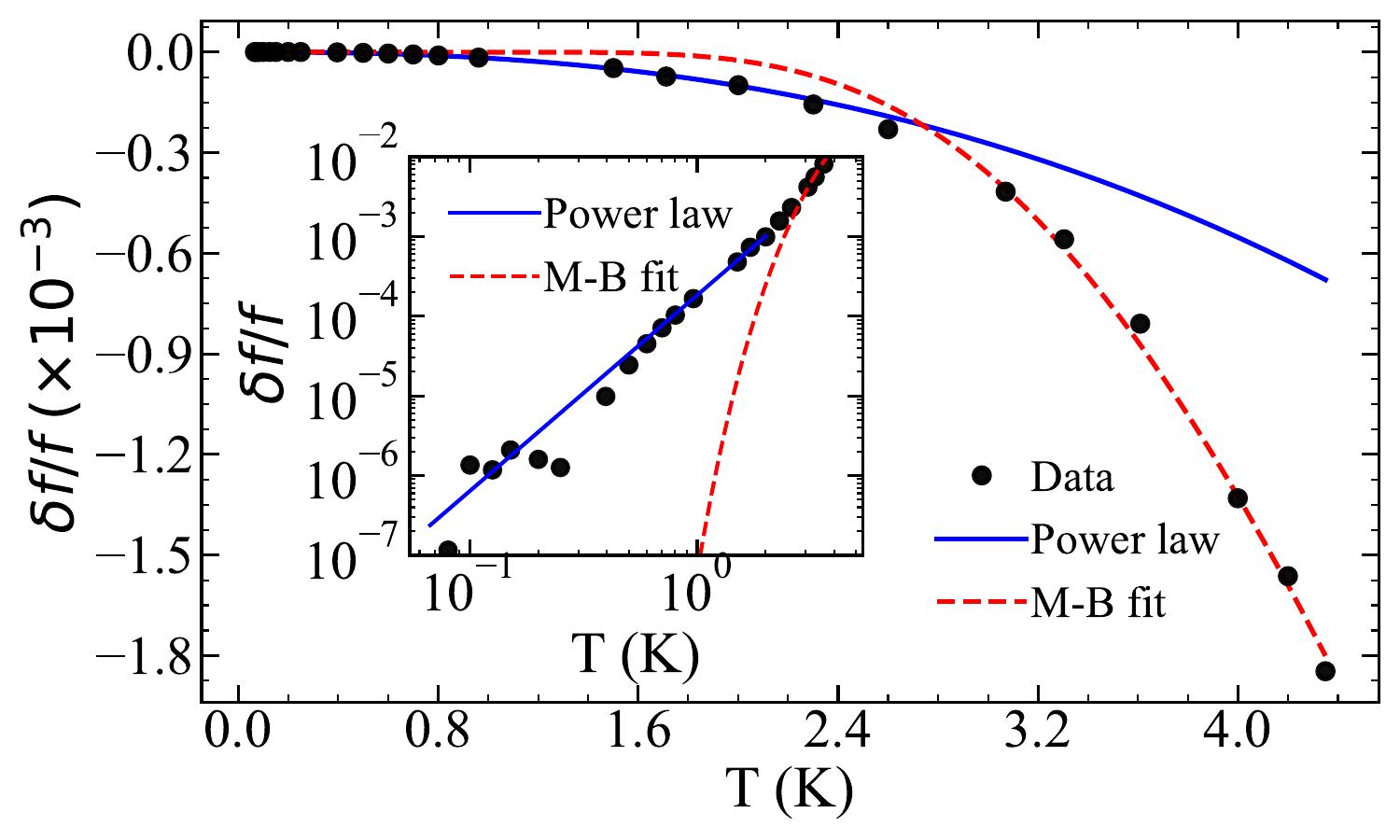}
        \caption{}
    \end{subfigure}
    \hfill
    \begin{subfigure}[t]{0.48\textwidth}
        \centering
        \includegraphics[width=\linewidth]{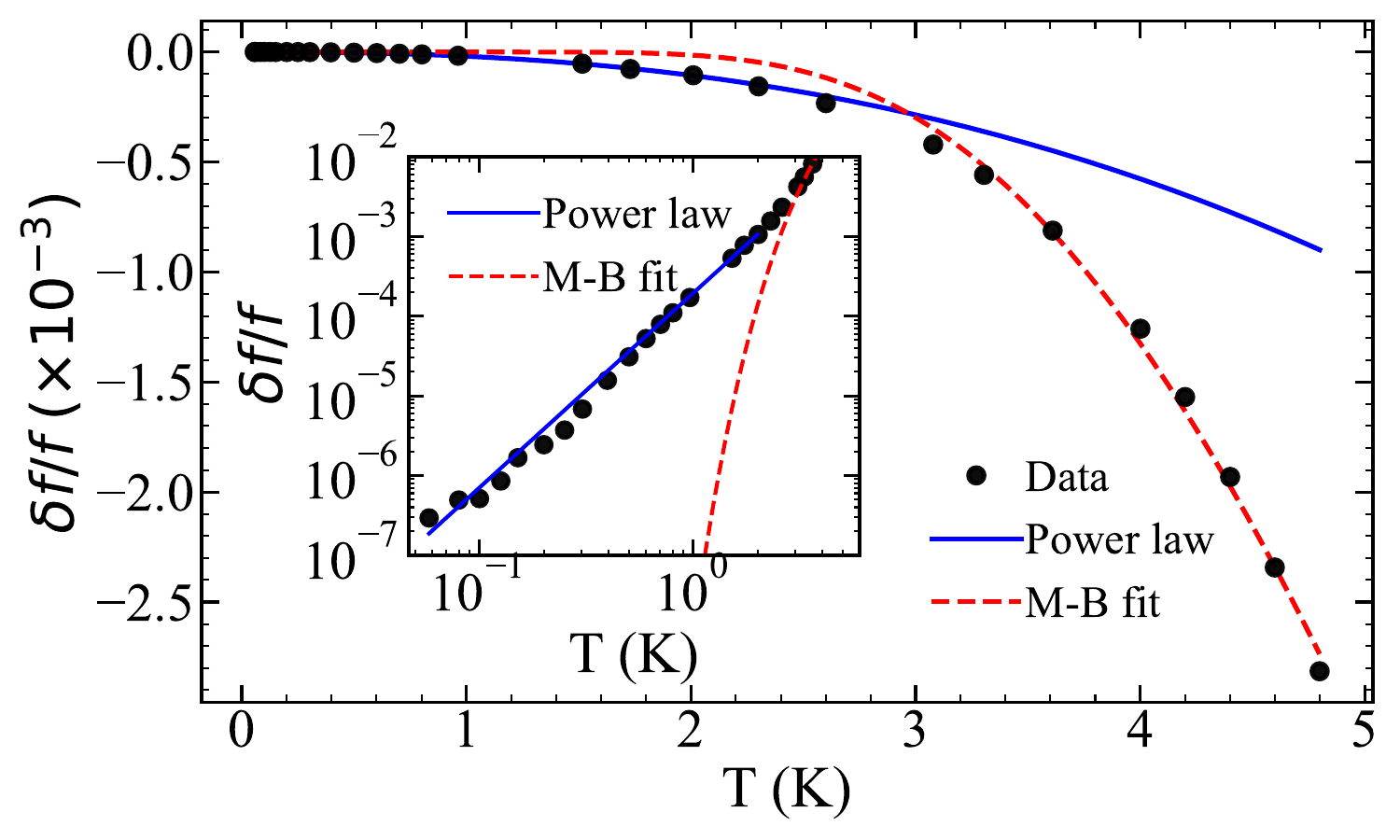}
        \caption{}
    \end{subfigure}

    \vspace{0.5em}

    \begin{subfigure}[t]{0.48\textwidth}
        \centering
        \includegraphics[width=\linewidth]{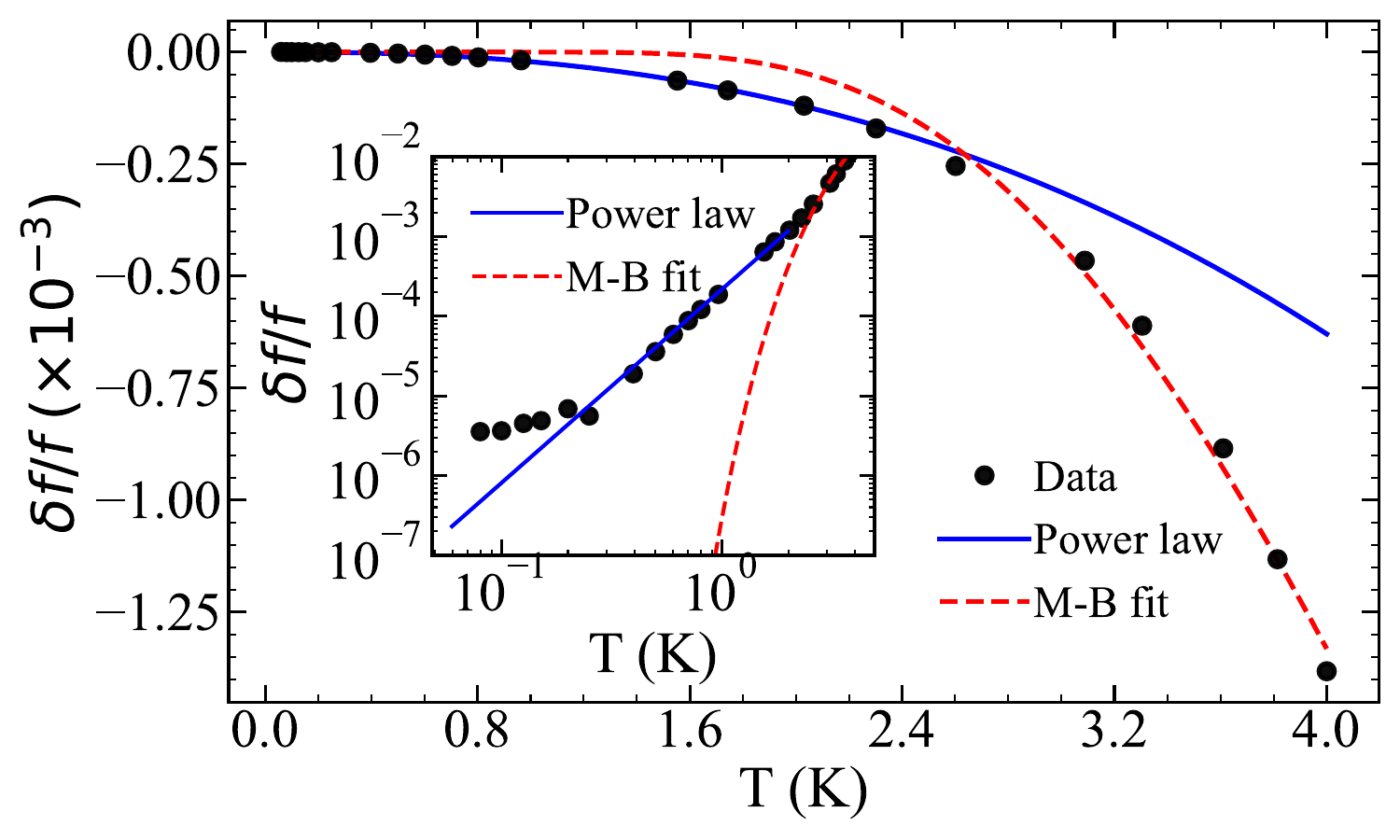}
        \caption{}
    \end{subfigure}
    \caption{\justifying {Temperature-dependent fractional frequency shift for 3~nm NbN resonators from the sister sample. 
    Panels (a), (b), and (c) correspond to measurements at different resonance frequencies 
    (3.6, 5.4, and 6.2~GHz, respectively). The data (black circles) are fit to both the 
    power-law model (blue lines) and the Mattis-Bardeen (M-B) model (red dashed lines). 
    The fitted power-law exponents are $b = 2.4564$, $b = 2.4417$, and $b = 2.4279$ 
    for the 3.6, 5.4, and 6.2~GHz modes, respectively.}}
    \label{fig:3nm_freq_shift_panel}
\end{figure}

The resonance modes shown by the sister samples of 2.8 nm NbN films are at 3.6~GHz, 5.4~GHz and 6.2~GHz. Figure~\ref{fig:3nm_freq_shift_panel} shows the temperature-dependent fractional frequency shift $\Delta f/f_0$. Each mode exhibits consistent power-law behavior $\Delta f/f_0 \propto T^b$ with exponents $b \approx 2.6$. Deviations from the fit at the lowest temperatures ($T < 100$~mK) are attributed to increased measurement noise near the base temperature of the dilution refrigerator.
 
\begin{figure}[htbp]
    \centering
    \begin{subfigure}[t]{0.48\textwidth}
        \centering
        \includegraphics[width=\linewidth]{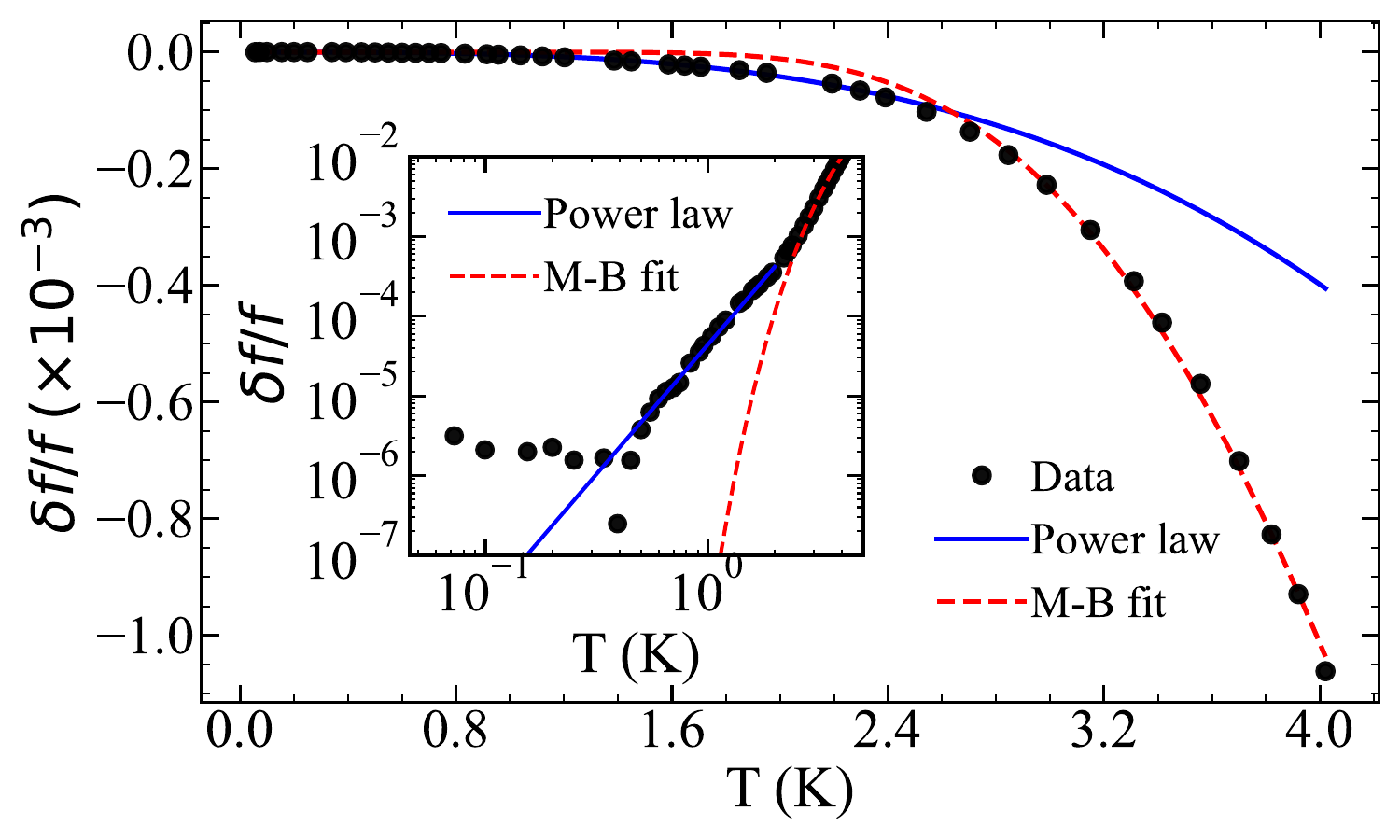}
        \caption{5~nm}
        \label{fig:5nm_freq_shift}
    \end{subfigure}
    \hfill
    \begin{subfigure}[t]{0.48\textwidth}
        \centering
        \includegraphics[width=\linewidth]{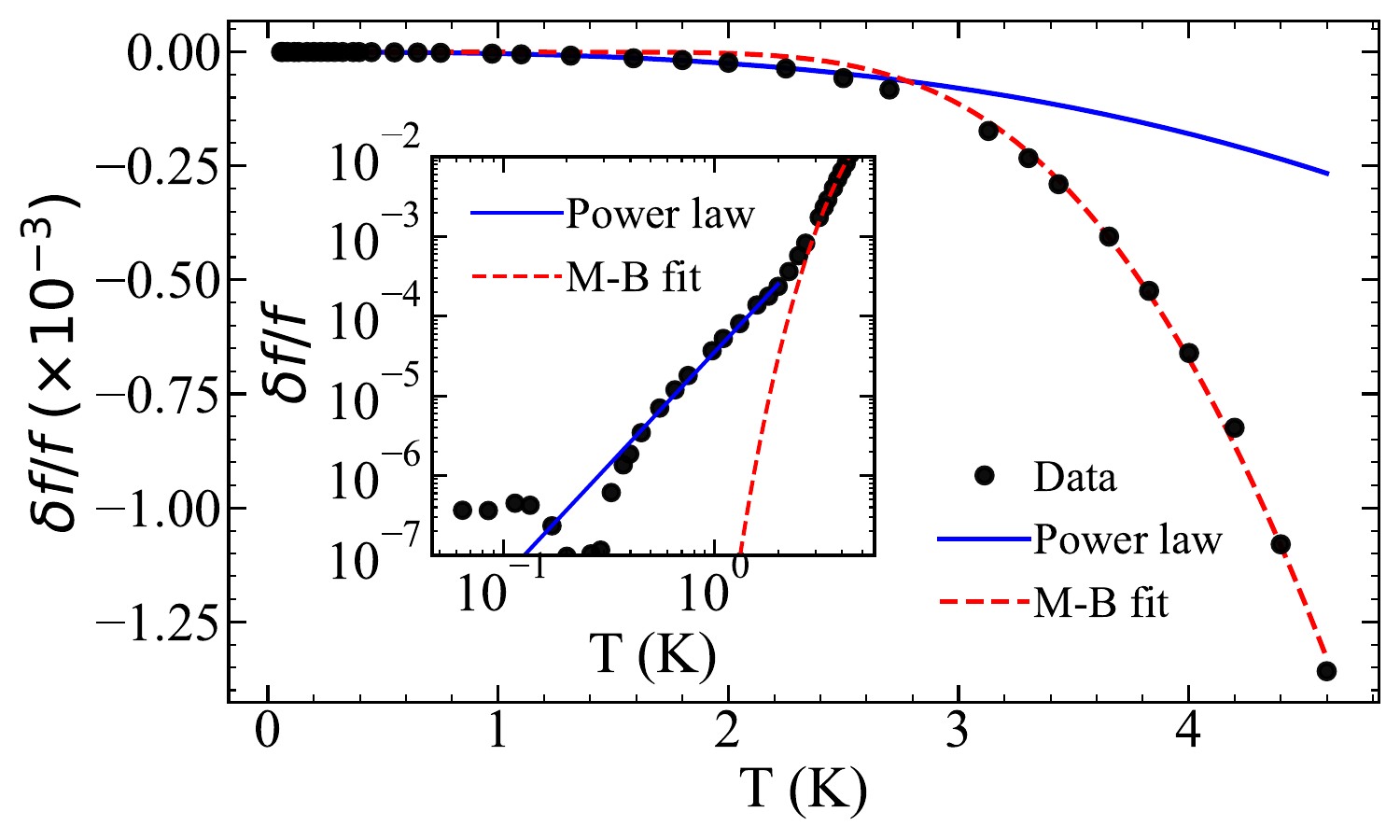}
        \caption{7~nm}
        \label{fig:7nm_freq_shift}
    \end{subfigure}
    \caption{\justifying {Temperature-dependent fractional frequency shift for an additional mode from the 
    (a)~5~nm and (b)~7~nm NbN samples. The data (black circles) are fit to both the 
    power-law model (blue line) and the Mattis-Bardeen (M-B) model (red dashed line). 
    The fitted power-law exponents are $b = 3.2490$ for the 5~nm sample and 
    $b = 2.8314$ for the 7~nm sample.}}
    \label{fig:5nm_7nm_freq_shift}
\end{figure}

\begin{figure}[htbp]
    \centering
    \includegraphics[width=0.49\textwidth]{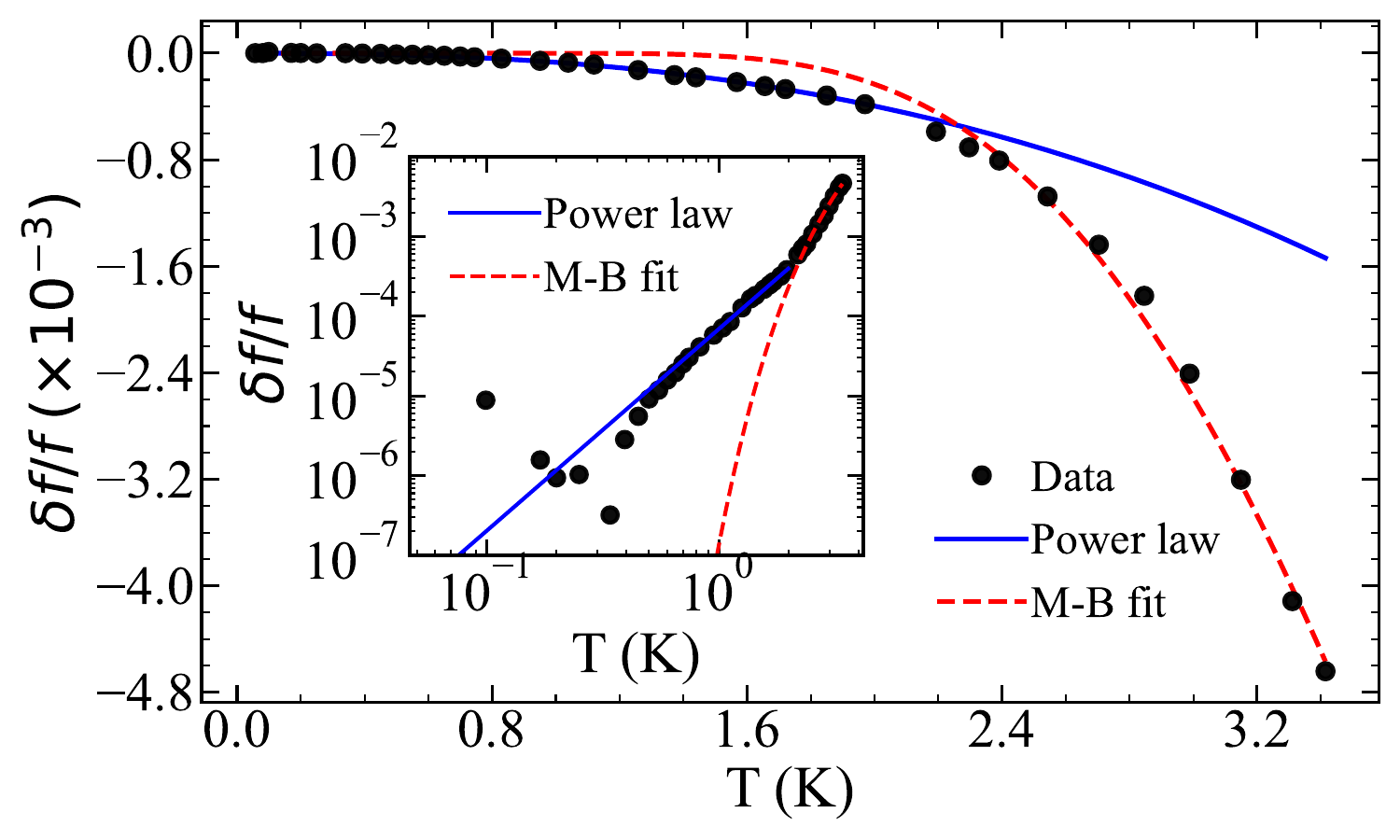}
    \caption{ \justifying {Temperature-dependent fractional frequency shift for another mode from the 
    25~nm NbN sample. The data (black circles) are fit to both the power-law model 
    (blue line) and the Mattis-Bardeen (M-B) model (red dashed line). 
    The fitted power-law exponent is $b = 2.5296$.
    }}
    \label{fig:25nm_freq_shift}
\end{figure}

\FloatBarrier

\section{Loss Mechanism Analysis}
\begin{figure}[htbp]  
    \centering

    \begin{subfigure}[t]{0.48\textwidth}
        \centering
        \includegraphics[width=\linewidth]{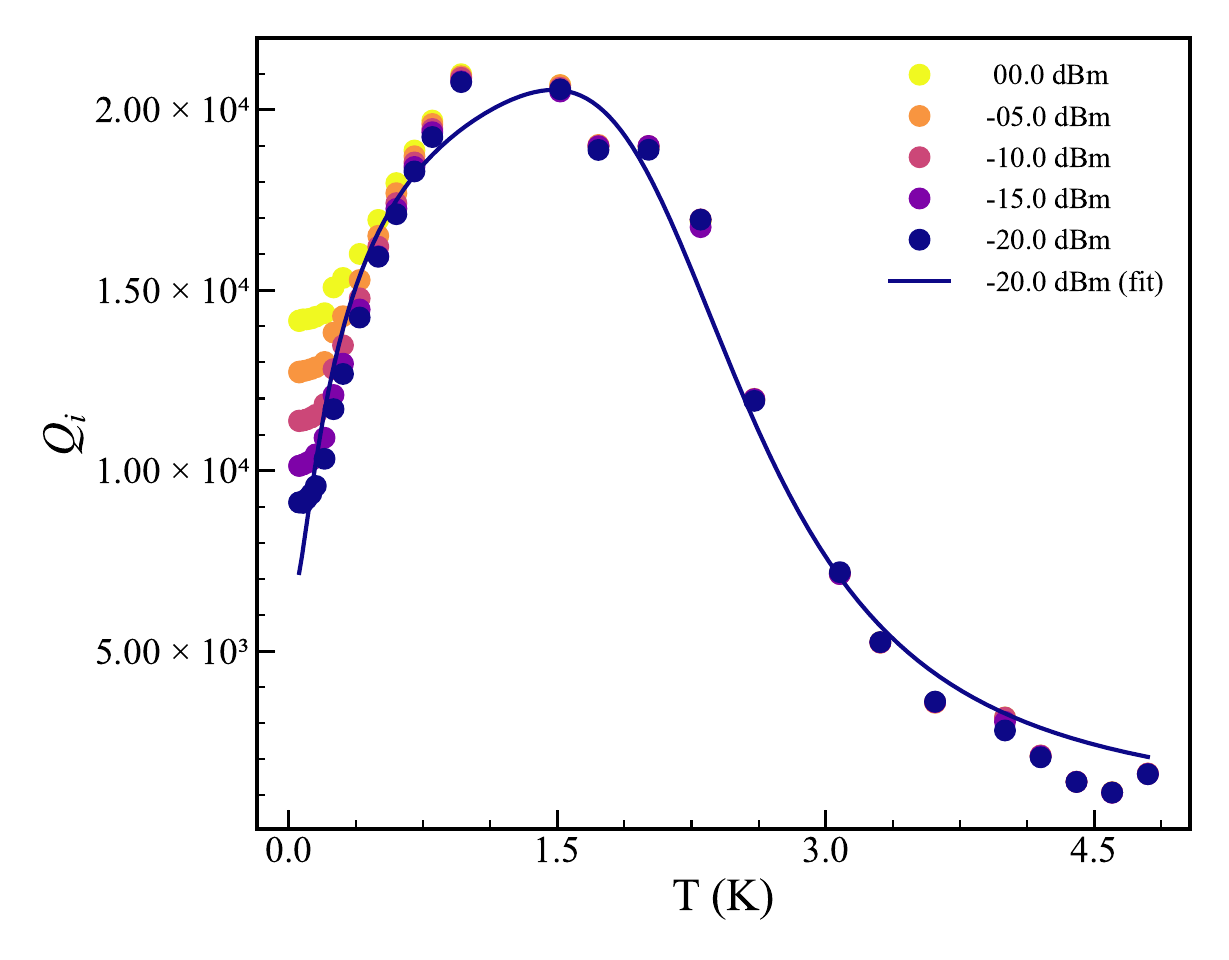}
        \caption{5.4 GHz mode}
    \end{subfigure}
    \hfill
    \begin{subfigure}[t]{0.48\textwidth}
        \centering
        \includegraphics[width=\linewidth]{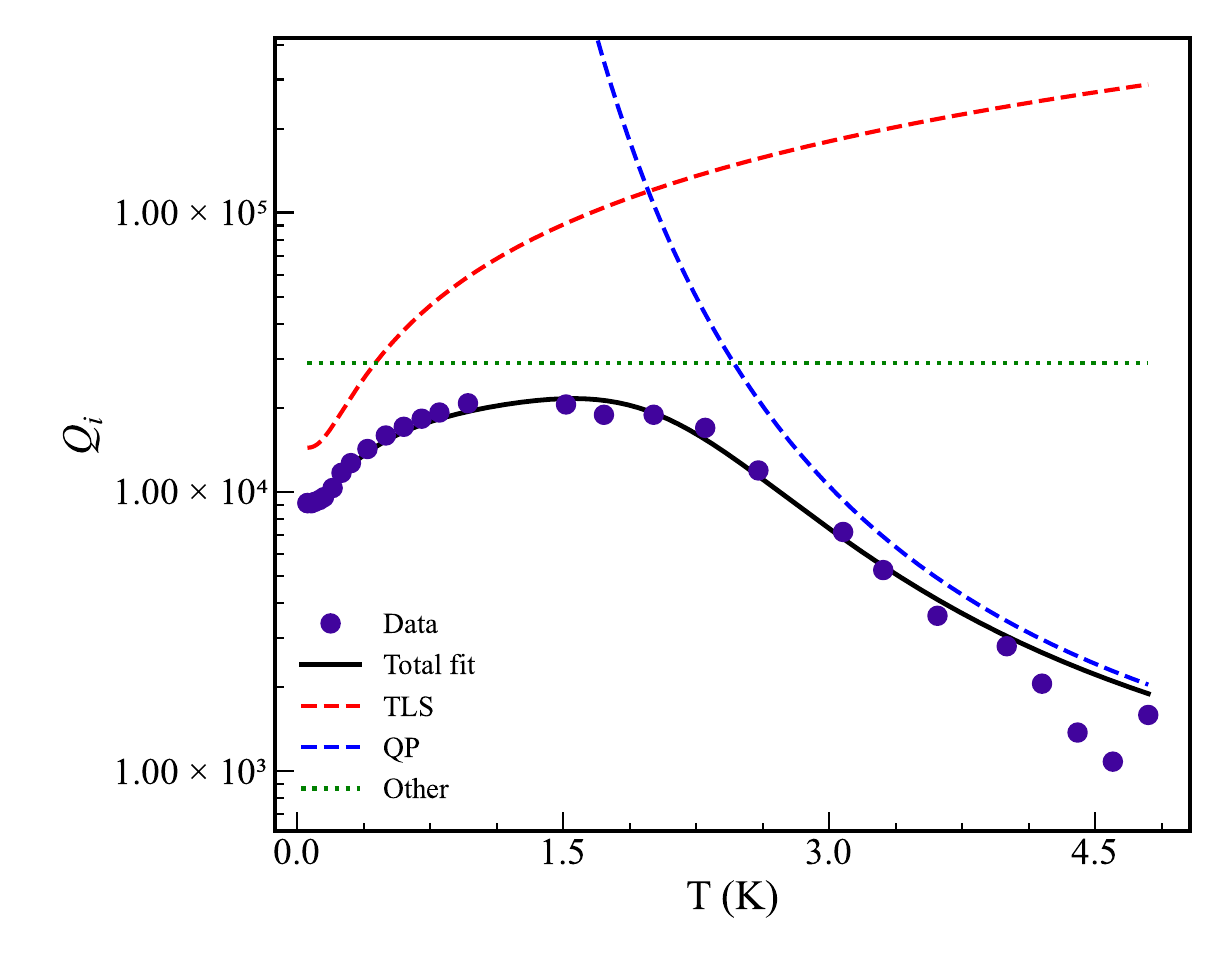}
        \caption{5.4 GHz fit}
    \end{subfigure}

    \vspace{0.6em}

    \begin{subfigure}[t]{0.48\textwidth}
        \centering
        \includegraphics[width=\linewidth]{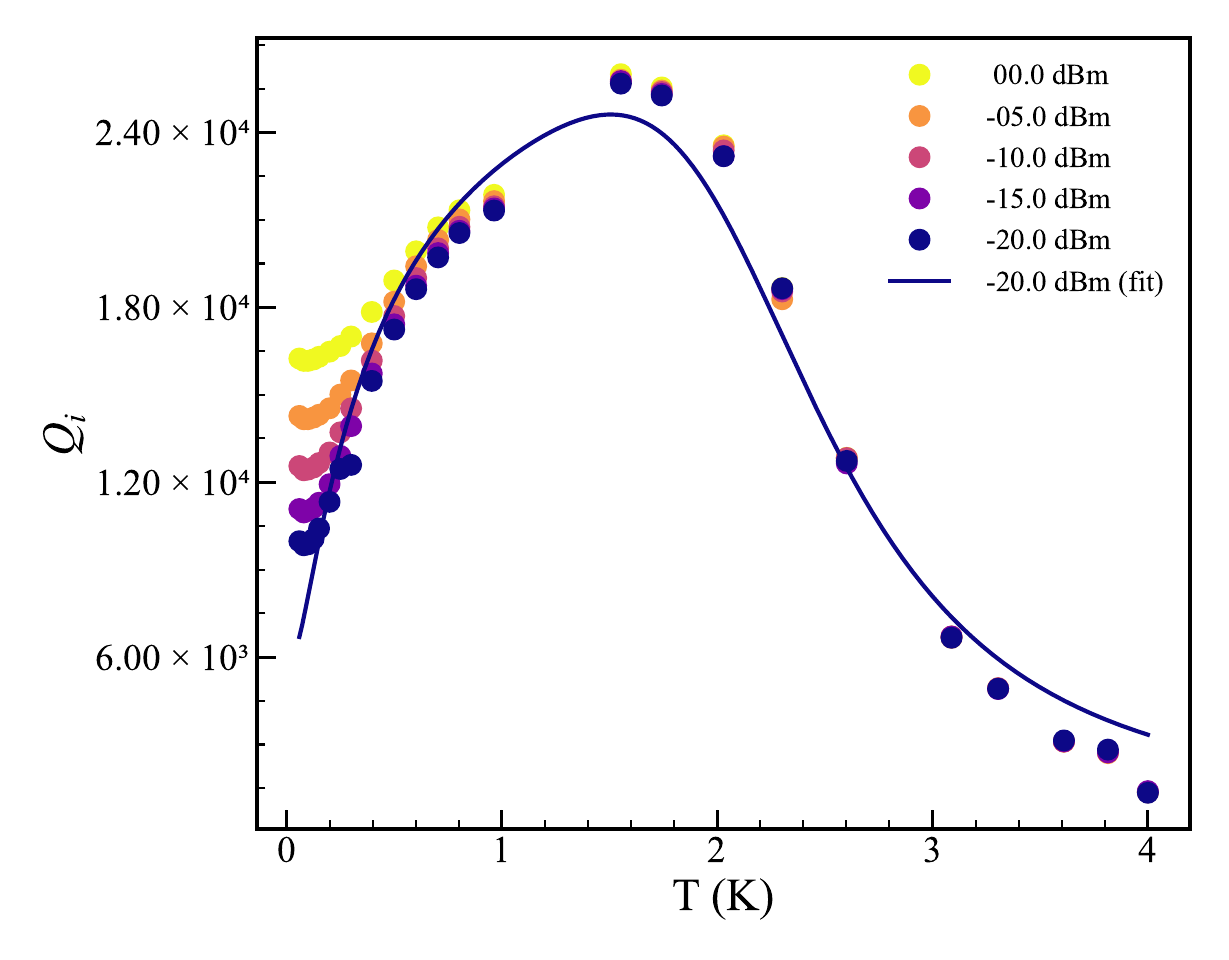}
        \caption{6.2 GHz mode}
    \end{subfigure}
    \hfill
    \begin{subfigure}[t]{0.48\textwidth}
        \centering
        \includegraphics[width=\linewidth]{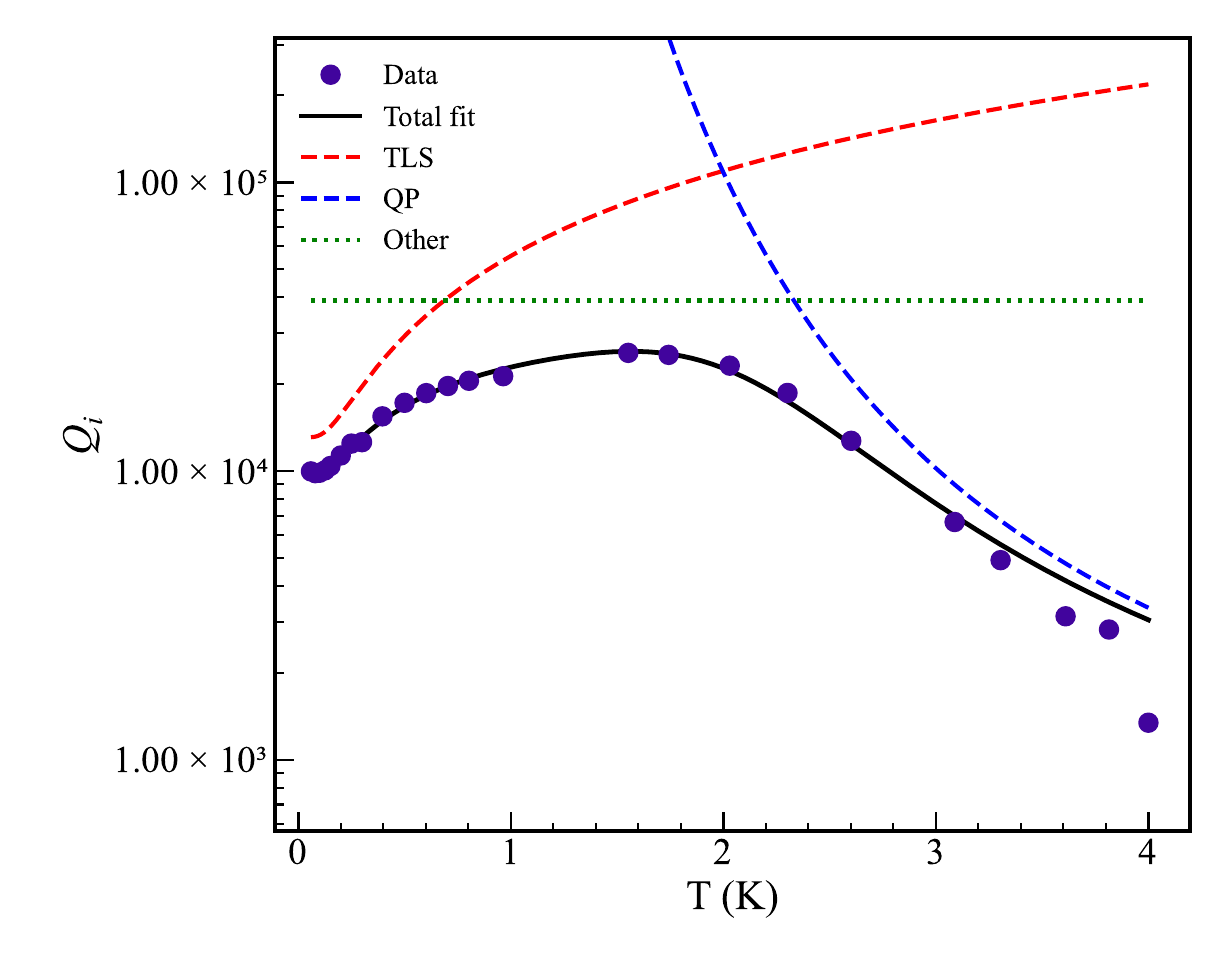}
        \caption{6.2 GHz fit}
    \end{subfigure}

    \caption{\justifying{
    Temperature dependence of the internal quality factor for the 2.8 nm NbN resonator. The panels correspond to resonance frequencies of 5.4~GHz and 6.2~GHz. The solid lines are fits to the combined TLS and quasiparticle loss model [Eq.~(\ref{eq:Qint})]. TLS losses dominate the low-temperature regime, while quasiparticle losses become dominant at higher temperatures. The fits, performed using the lowest-power dataset, yield $Q_{\mathrm{TLS},0} \approx 5.99 \times 10^3$ for the 5.4~GHz mode and $Q_{\mathrm{TLS},0} \approx 1.31 \times 10^4$ for the 6.2~GHz mode.}}
  
     \label{fig:3nm_Q_panel}
\end{figure}

We model the internal loss of the resonators as the sum of three contributions: 
two-level-system (TLS) loss \cite{Martinis2005, Gao2008Thesis, Gao2008APL}, quasiparticle loss 
\cite{Mattis1958, Zmuidzinas2012}, and a temperature-independent residual term. Because loss rates from independent channels add in the inverse quality factor, we extract each contribution by fitting the measured \(Q_i(T)\) to this combined model, exploiting the distinct temperature dependence of the two intrinsic channels.
TLS loss arises from dielectric two-level defects in the amorphous oxides and interfaces surrounding the resonator; it dominates at low temperature and is progressively saturated as temperature or drive power increases. Quasiparticle loss originates from thermally excited unpaired electrons above the superconducting gap, a population that grows as \(\exp[-\Delta(0)/k_B T]\) and therefore dominates at higher temperatures. The residual term collects all temperature-independent loss channels not captured by these two mechanisms- radiation into the substrate or free space, residual vortex motion, and fabrication imperfections and sets the saturation floor of \(Q_i\) once the TLS and quasiparticle contributions become negligible.
 \begin{equation}
    \frac{1}{Q_{\text{int}}} = \frac{1}{Q_{\text{TLS}}(T)} 
    + \frac{1}{Q_{\text{QP}}(T)} 
    + \frac{1}{Q_{\text{other}}}.
    \label{eq:Qint}
\end{equation}
The temperature dependence of the TLS contribution is modeled as \cite{Gao2008Thesis}
\begin{equation}
    Q_{\text{TLS}}(T) = \frac{Q_{\text{TLS},0}}{\tanh\left(\frac{\hbar \omega}{2 k_B T}\right)},
    \label{eq:QTLS}
\end{equation}
which captures the fact that TLS are most effective at low temperatures 
($k_B T \ll \hbar \omega$), while at higher temperatures they gradually saturate 
and contribute less to the total loss.\\
The quasiparticle losses are described by the Mattis-Bardeen theory~\cite{Mattis1958}:
\begin{equation}
    Q_{\text{QP}}(T) = Q_{\text{QP},0} \, 
    \frac{e^{\Delta_0/k_B T}}
    {\sinh\left(\frac{\hbar \omega}{2 k_B T}\right) 
    K_0\left(\frac{\hbar \omega}{2 k_B T}\right)},
    \label{eq:QQP}
\end{equation}

where $\Delta_0$ is the superconducting gap energy and $K_0$ is the modified 
Bessel function of the second kind. This form reflects the exponential suppression 
of quasiparticles at millikelvin temperatures and their increasing density when the 
temperature approaches the superconducting gap.
The experimental data for the internal quality factor were fitted to 
Eq.~\eqref{eq:Qint} using Eqs.~\eqref{eq:QTLS} and~\eqref{eq:QQP} 
with $Q_{\text{TLS},0}$, $Q_{\text{QP},0}$, $\Delta_0$, and $Q_{\text{other}}$ 
treated as fitting parameters. This allowed us to disentangle 
the relative importance of TLS, quasiparticle, and residual losses 
in limiting resonator performance.\\
The internal quality factor $Q_i$ of the resonators was measured as a function of temperature for several input microwave power levels ranging from $-20$~dBm to $0$~dBm. 
Figure~\ref{fig:3nm_Q_panel} shows the internal quality factor $Q_i$ as a function of temperature for various input powers ranging from $-20$~dBm to $0$~dBm for the 2.8 nm film NbN resonator. We fit the TLS plus quasiparticle model to the lowest-power data ($-20$~dBm). At low powers, the resonator operates in the linear regime where TLS losses are unsaturated and the fitted parameters ($Q_{\text{TLS},0} \approx 5.99 \times 10^3$ for the 5.4~GHz mode at $-20$~dBm) are consistent with literature values for similar systems \cite{Jin2025, Niepce2019}.

\begin{figure}[htbp]
    \centering

    \begin{subfigure}[t]{0.70\textwidth}
        \centering
        \includegraphics[width=0.76\linewidth]{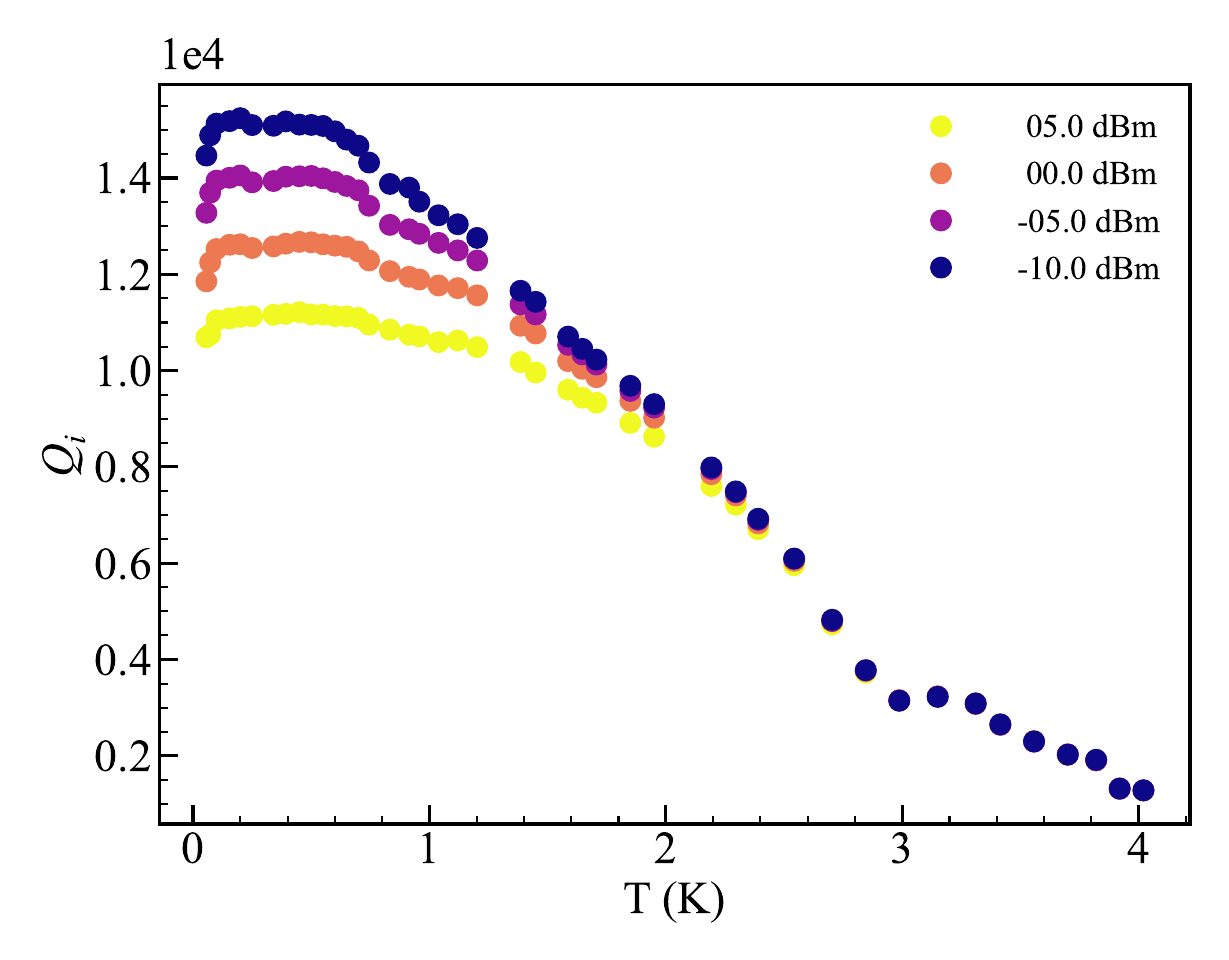}
        \caption{5 nm mode}
    \end{subfigure}

    \caption{\justifying{
    Temperature-dependent internal quality factor for the 5 nm NbN resonator. Unlike other devices, the measured $Q_i(T)$ data exhibit irregular temperature dependence and do not follow the expected trend consistently. Therefore, no fit to the TLS and quasiparticle model is presented for this device.
    }}
\end{figure}

\begin{figure}[htbp]
    \centering

    \begin{subfigure}[t]{0.48\textwidth}
        \centering
        \includegraphics[width=1.05\linewidth]{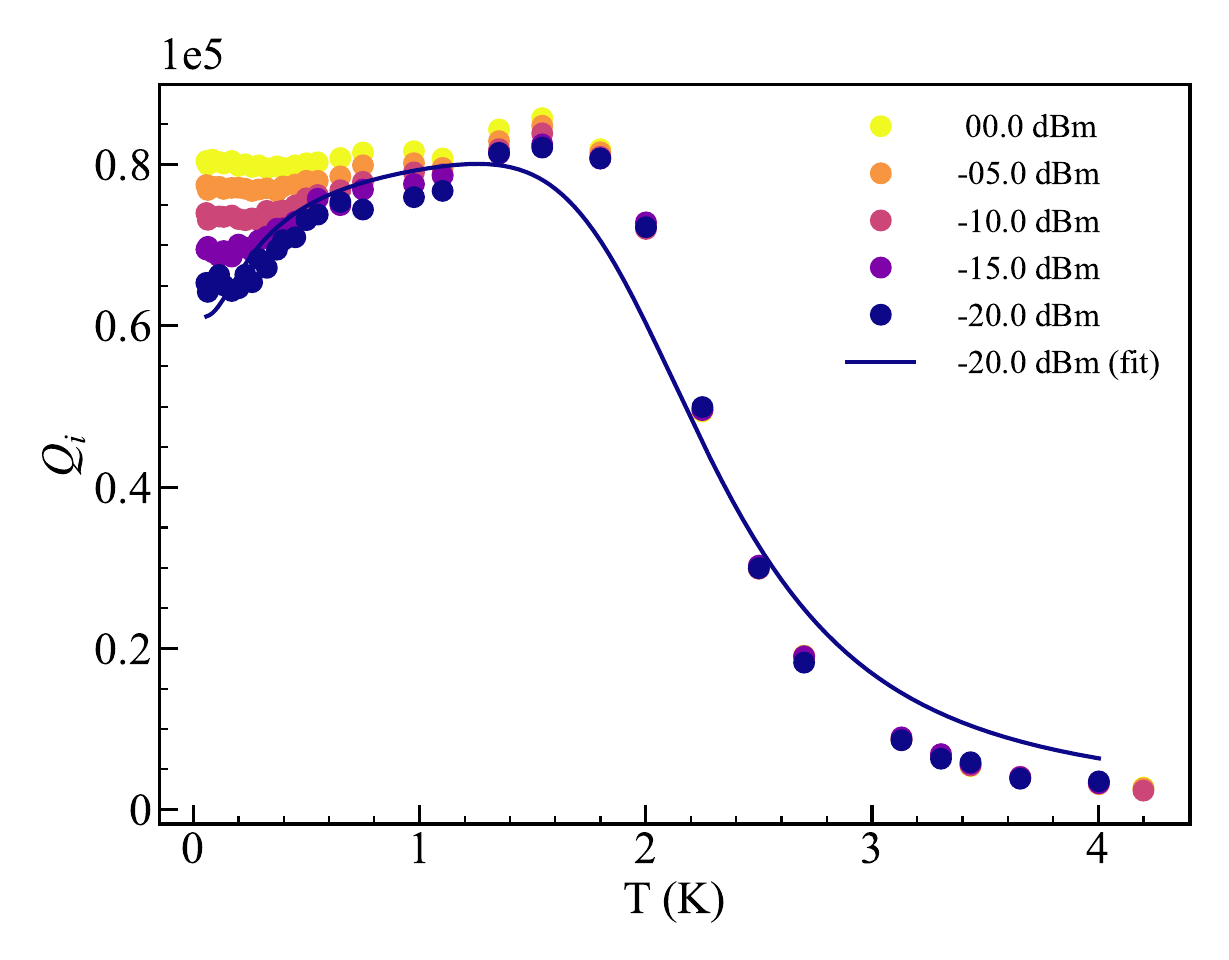}
        \caption{7 nm mode}
    \end{subfigure}
    \hfill
    \begin{subfigure}[t]{0.48\textwidth}
        \centering
        \includegraphics[width=\linewidth]{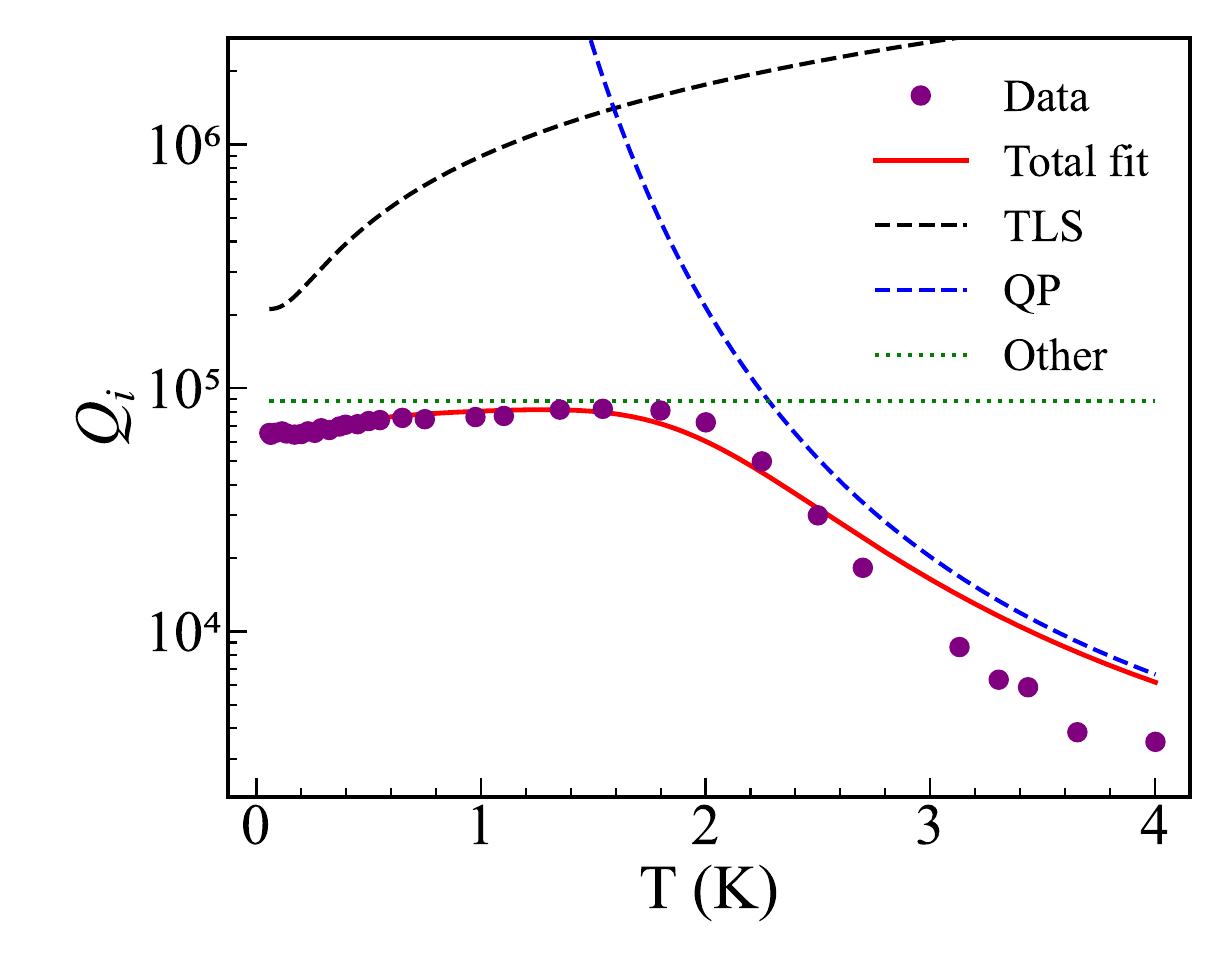}
        \caption{7 nm fit}
    \end{subfigure}

    \caption{\justifying{
    Temperature dependence of the internal quality factor for the 7 nm NbN resonator. The solid line is a fit to the combined TLS and quasiparticle loss model [Eq.~(\ref{eq:Qint})], performed using the lowest-power dataset, yielding $Q_{\mathrm{TLS},0} \approx 4.53 \times 10^5$.}}
    
\end{figure}

\begin{figure}[htbp]
    \centering

    \begin{subfigure}[t]{0.48\textwidth}
        \centering
        \includegraphics[width=1.07\linewidth]{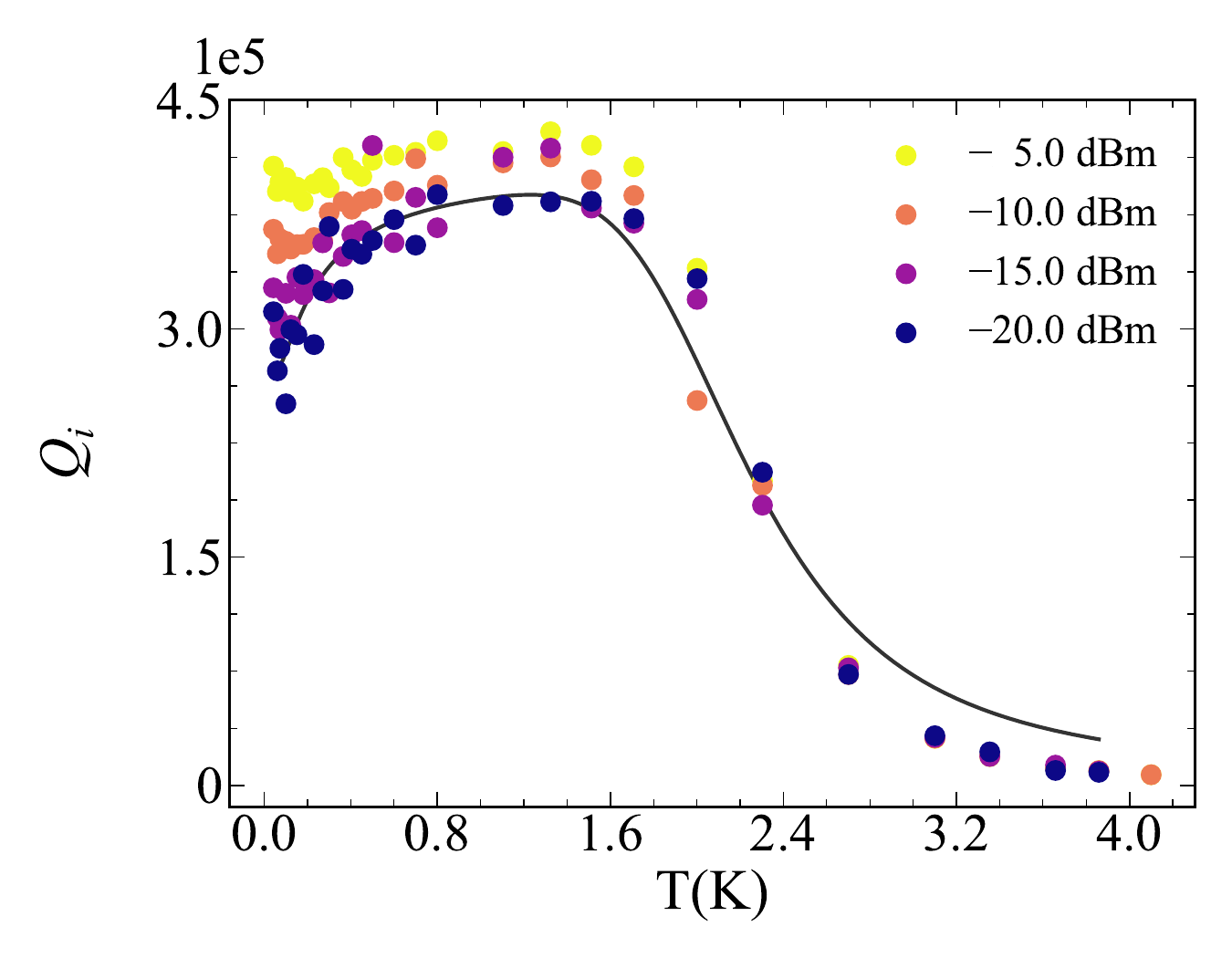}
        \caption{25 nm mode}
    \end{subfigure}
    \hfill
    \begin{subfigure}[t]{0.48\textwidth}
        \centering
        \includegraphics[width=\linewidth]{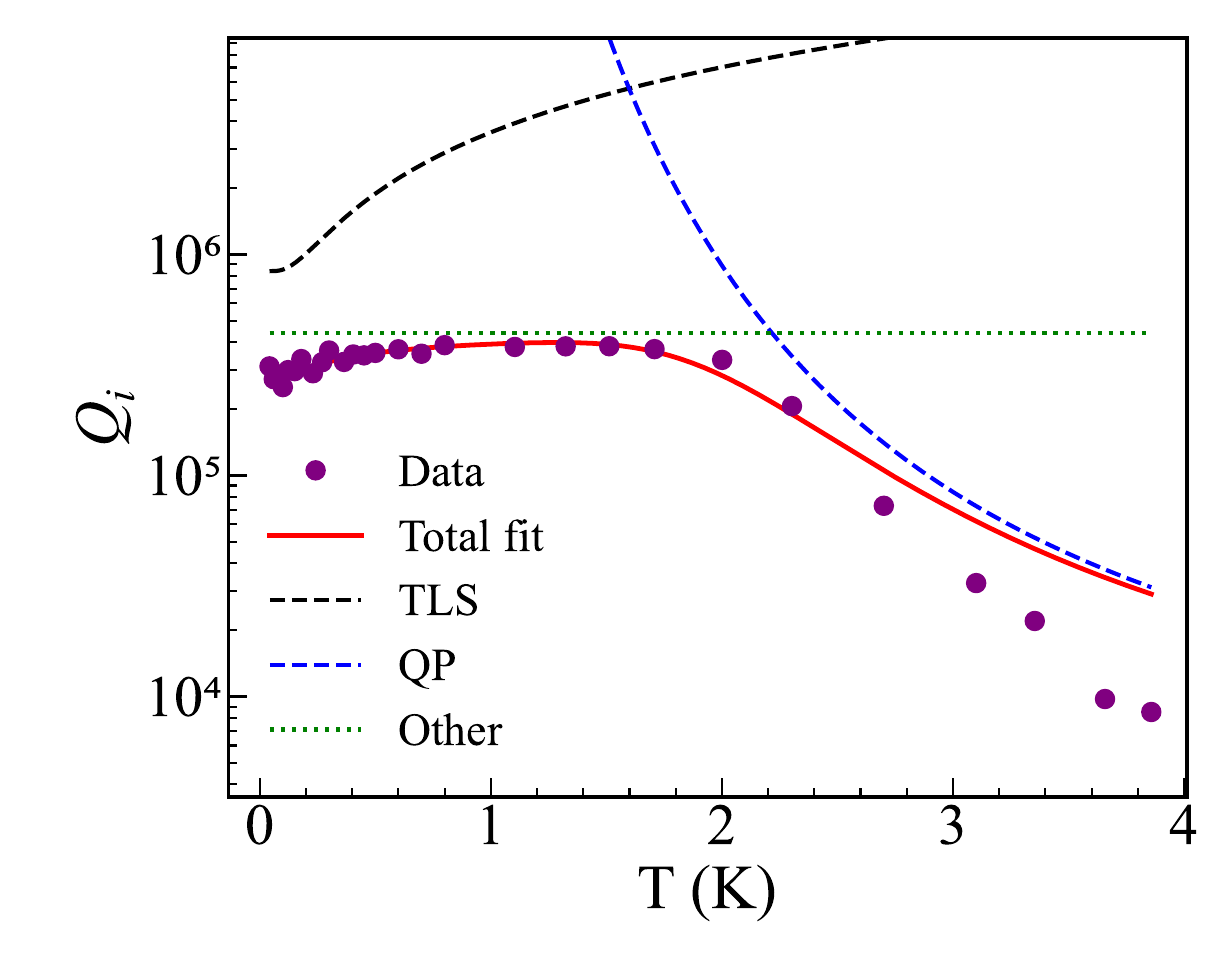}
        \caption{25 nm fit}
    \end{subfigure}

    \caption{\justifying {
    Temperature-dependent internal quality factor of the 25 nm NbN resonator. The solid line shows a fit to the combined TLS and quasiparticle model [Eq.~(\ref{eq:Qint})] using the lowest-power data, yielding $Q_{\mathrm{TLS},0} \approx 2.85 \times 10^6$.}}
\end{figure}
\begin{figure}[htbp]
    \centering
    \includegraphics[width=0.6\linewidth]{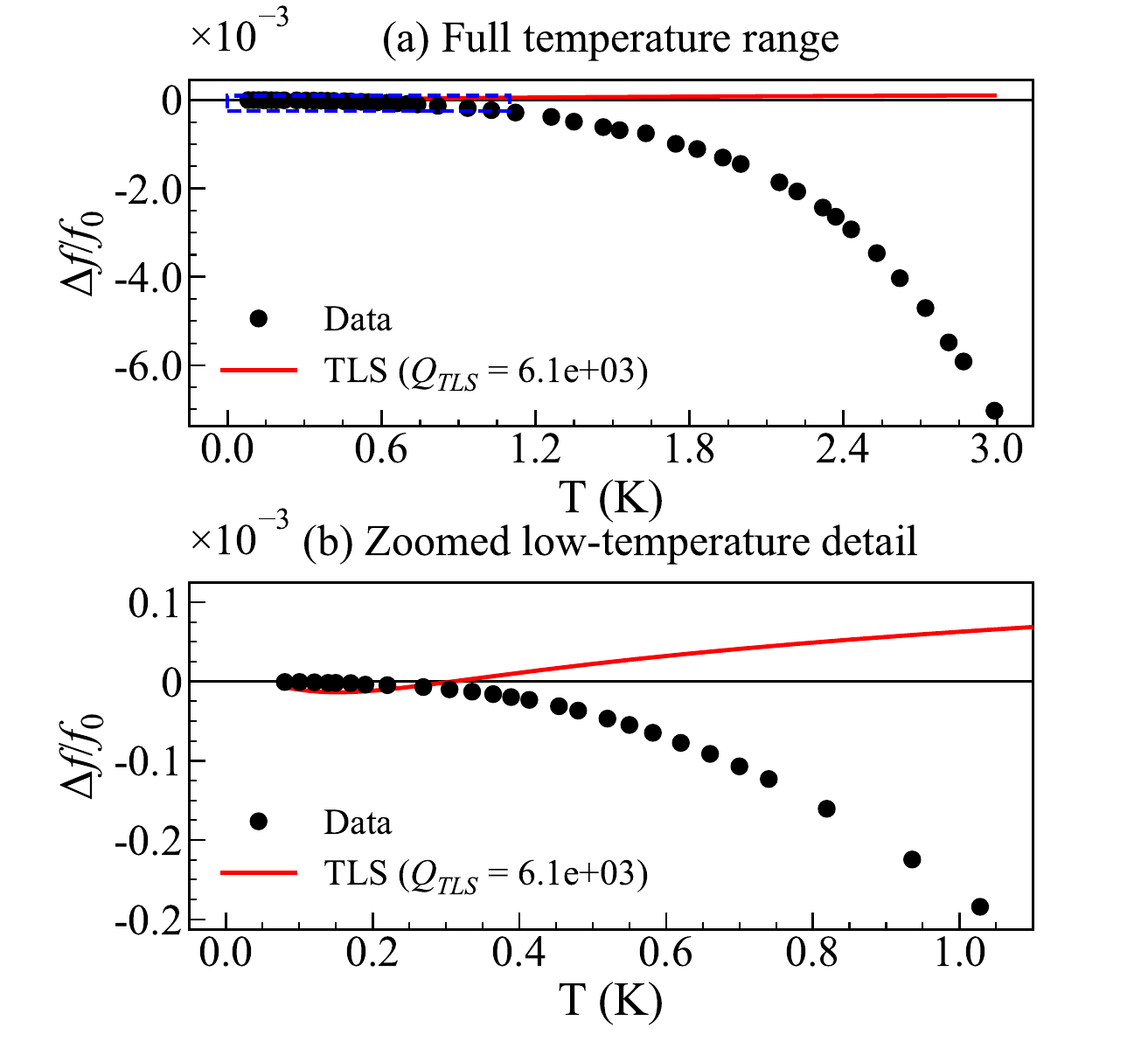}
    \caption{\justifying {
    Expected TLS dispersive frequency shift (red line) calculated using 
    $Q_{\text{TLS},0} = 6.1 \times 10^3$ alongside the measured frequency shift 
    data (black circles) for the 2.8~nm NbN resonator. The predicted TLS shift is 
    significantly smaller than the kinetic inductance dominated response, explaining 
    why it is not seen in the data.
    }}
    \label{fig:TLS_freq_shift}
\end{figure}

In addition to their dissipative contribution, TLS are also expected to produce a dispersive frequency shift through their coupling to the reactive part of the resonator response. Using the $Q_{\text{TLS},0}$ extracted from the $Q_i$ fits, the expected TLS-induced frequency shift can be estimated as $\delta f / f_0 \sim 1/(2\pi Q_{\text{TLS},0}) \cdot \tanh(\hbar\omega / 2k_BT)$, which yields values on the order of $10^{-5}$ across the measured temperature range. As shown in Figure~\ref{fig:TLS_freq_shift}, this predicted shift is significantly smaller in magnitude than the frequency shift due to the temperature-dependent kinetic inductance of the 2.8~nm NbN film, which dominates the dispersive response. Consequently, the TLS dispersive contribution is not resolved in our frequency shift data, consistent with the expectation that kinetic inductance effects overwhelm the dielectric 
response in the ultrathin-film limit. The observed behavior is therefore consistent with the TLS loss picture established through the $Q_i$ analysis.

